\documentclass[12pt,reprint,aps,groupedaddress,nofootinbib,prd,twocolumn]{revtex4-2}

\usepackage[english]{babel}
\usepackage[utf8]{inputenc}
\usepackage{amsmath}
\usepackage{mathbbol}
\usepackage{amssymb}
\usepackage{bbold}
\usepackage{graphicx,amsfonts}
\usepackage{epsfig}
\usepackage[colorlinks=true,
linkcolor=blue,
urlcolor=blue,
citecolor=blue]{hyperref}
\usepackage{bm}
\usepackage{mathrsfs}
\usepackage{enumerate}
\usepackage{amsthm}
\usepackage{bbm}
\usepackage{comment}
\usepackage{physics}
\usepackage{url}
\usepackage[usenames,dvipsnames]{xcolor}

\newcommand{\be}{\begin{equation}}
\newcommand{\ee}{\end{equation}}
\newcommand{\bea}{\begin{eqnarray}}
\newcommand{\eea}{\end{eqnarray}}

\def\scri{\mathscr{I}}

\begin{document}

\title{Structural aspects of the anti-de Sitter black hole pseudospectrum}

\author{Valentin Boyanov$^{1}$, Vitor Cardoso$^{1,2}$, Kyriakos Destounis$^{3,4}$, Jos\'e Luis Jaramillo$^{5}$ and Rodrigo Panosso Macedo$^{2}$} 
\affiliation{${^1}$CENTRA, Departamento de F\'{\i}sica, Instituto Superior T\'ecnico -- IST, Universidade de Lisboa -- UL, Avenida Rovisco Pais 1, 1049 Lisboa, Portugal}
\affiliation{${^2}$ Niels Bohr International Academy, Niels Bohr Institute, Blegdamsvej 17, 2100 Copenhagen, Denmark}
\affiliation{$^3$Dipartimento di Fisica, Sapienza Università di Roma, Piazzale Aldo Moro 5, 00185, Roma, Italy}
\affiliation{$^4$INFN, Sezione di Roma, Piazzale Aldo Moro 2, 00185, Roma, Italy}
\affiliation{${^5}$Institut de Math\'ematiques de Bourgogne (IMB), UMR 5584, CNRS, Universit\'e de Bourgogne, F-21000 Dijon, France}

\begin{abstract}
Black holes in anti-de Sitter spacetime provide an important testing ground for both gravitational and field-theoretic phenomena. In particular, the study of perturbations can be useful to further our understanding regarding certain physical processes, such as superradiance, or the dynamics of strongly coupled conformal field theories through the holographic principle. In this work we continue our systematic study of the ultraviolet instabilities of black-hole quasinormal modes, built on the characterization of the latter as eigenvalues of a (spectrally unstable) non-selfadjoint operator and using the pseudospectrum as a main analysis tool, extending our previous studies in the asymptotically flat setting to Anti-de Sitter asymptotics. Very importantly, this step provides a singularly well-suited probe into some of the key structural aspects of the pseudospectrum. This is a consequence of the specific features of the Schwarzschild-anti-de Sitter geometry, together with the existence of a sound characterization by Warnick of quasinormal modes as eigenvalues, that is still absent in asymptotic flatness. This work focuses on such structural aspects, with an emphasis on the convergence issues of the pseudospectrum and, in particular, the comparison between the hyperboloidal and null slicing cases. As a physical by-product of this structural analysis we assess, in particular, the spectral stability of purely imaginary ``hydrodynamic" modes, which appear for axial gravitational perturbations, that become dominant when the black-hole horizon is larger than the anti-de Sitter radius. We find that their spectral stability, under perturbations, depends on how close they are to the real axis, or conversely how distant they are from the first oscillatory overtone.
\end{abstract}

\maketitle

\tableofcontents

\section{Introduction \label{s:section}}
Quasinormal mode (QNM) frequencies of black hole (BH) spacetimes, namely the complex frequencies capturing the linear response to external perturbations and encoding intrinsic geometric information of the background, play a major role in different aspects of gravitational physics \cite{Kokkotas:1999bd,Nollert:1999ji,Berti:2009kk,Konoplya:2011qq,zworski2017mathematical}, from fundamental stability results in mathematical relativity,
applications to the late-time ringdown phase of dynamical BH spacetimes in gravitational-wave (GW) astronomy, high-energy astrophysical phenomena, such as superradiance and superradiant instabilities \cite{Brito:2015oca,Destounis:2019hca,Mascher:2022pku,Destounis:2022rpk}, or in the bulk-boundary duality in the anti-de Sitter/conformal field theory (AdS/CFT) setting
\cite{Maldacena:1997re,Hubeny:2014bla}. Therefore, the study of their structural stability aspects is of major importance. Recent studies have put a focus on the QNM spectral stability under small perturbations of the environment \cite{Barausse:2014tra,Cardoso:2021wlq,Cardoso:2022whc,Courty:2023rxk}, raising fundamental questions that remain open. In this work we focus on this problem dwelling in the setting of asymptotically AdS spacetimes, since they provide a particularly well controlled mathematical environment for the proof-of-principle study that we aim at here.

\subsection{QNM spectral instability in BH spacetimes}
\label{s:QNM_spectral}
The sensitivity of BH QNM frequencies to small perturbations of the environment was early identified in the works by Nollert \& Price and  Aguirregabiria \&
Vishveshwara~\cite{Nollert:1996rf,Aguirregabiria:1996zy,Vishveshwara:1996jgz,Nollert:1998ys}, receiving since then some attention in the literature, mainly in the GW setting~\cite{Leung:1997was,Barausse:2014tra,Daghigh:2020jyk,Qian:2020cnz}.

Recently, an approach to the BH QNM stability problem 
based on the discussion of the eigenvalue stability in non-selfadjoint spectral problems has been proposed\footnote{The setting is more general than that of BH QNMs, actually extending to QNM stability in general linear wave equations, in particular under perturbations in generic potentials but also the permittivity function in the optical setting~\cite{alsheikh:tel-04116011}.}
in~\cite{Jaramillo:2020tuu,Jaramillo:2021tmt,alsheikh:tel-04116011,Gasperin2021}, in particular introducing the notion of
pseudospectrum~\cite{Trefethen:1993,Driscoll:1996,Trefethen:2005,Davie00,Davie07,Sjostrand2019} both into the QNM problem and in gravitational physics. This scheme has been then applied to a variety of BH settings, proving to be robust\footnote{A key aspect of this BH instability is its ultraviolet (high-wavenumber/low-regularity \cite{Zwors87,Berry82,BinZwo} and Refs. in \cite{Gasperin2021}) character, reflected in particular in the logarithmic universality of large overtones probing small scales that make that, as long as asymptotics/boundary conditions are leaky, their detailed form seems important for the qualitative QNM instability effect.} under various spacetime asymptotics (asymptotically flatness, de Sitter (dS) and AdS) and boundary conditions \cite{Jaramillo:2020tuu,Destounis:2021lum,Boyanov:2022ark,Sarkar:2023rhp,Arean:2023ejh,Cownden:2023dam}. Of particular importance in the present work is the systematic study in Arean et al.~\cite{Arean:2023ejh}, further extended\footnote{In the very late stage of the writing this manuscript, we became aware of the work \cite{Cownden:2023dam}. Results in \cite{Cownden:2023dam} and the ones here presented, in particular regarding the discussion of the pseudospectrum in a null slicing, are independent and complementary.} by Cownden et al. \cite{Cownden:2023dam}, that discuss the asymptotically AdS case providing, in particular, an excellent motivation for the study of the BH QNM spectral instability in the AdS/CFT context.
Beyond QNM spectral instability, other aspects of the pseudospectrum in the (non-selfadjoint) BH problem, namely transients and pseudo-resonances, have been explored in~\cite{Jaramillo:2022kuv,Boyanov:2022ark} (cf. also \cite{Cownden:2023dam}).

These BH QNM instability studies have prompted other related works exploring associated aspects~\cite{Cheung:2021bol,Berti:2022xfj,konoplya2209first,Torres:2023nqg}, not specifically using a non-selfadjoint spectral problem framework, but confirming the qualitative
instability picture, deepening in questions such as the instability of the fundamental QNM and opening new complementary problems. A recent review, focused on the GW BH spectroscopy problem, can be found in~\cite{Destounis:2023ruj}.

\subsubsection{Pseudospectrum and BH QNMs: open problems}
\label{s:Pseudospectrum_open_problems}
Given a selfadjoint operator $L$ (more generally a so-called `normal operator'~\cite{Trefethen:2005}, namely commuting with its adjoint $[L, L^\dagger]=0$), its spectrum $\sigma(L)$ encodes fundamental information of $L$ that is intrinsic to the operator, not depending on any other object. However, when the operator is non-normal ($[L, L^\dagger]\neq 0$) the spectrum $\sigma(L)$ is no longer necessarily the good notion to consider.

 In such a non-normal setting, a more robust notion that partially substitutes the spectrum, in particular regarding spectral stability and evolution aspects when $L$ is considered as an infinitesimal time generator, is the notion of the pseudospectrum (or more properly the $\epsilon$-pseudospectrum set) that we briefly revisit in Sec.~\ref{s:QNM_as_eigenvalues}. However, there is a price to pay: the pseudospectrum is not a notion intrinsic to $L$, but depends on the choice of a norm. This point has been addressed in detail in~\cite{Gasperin2021}. Specifically, the pseudospectrum sets can be characterized in terms of the norm of the resolvent $R_L(\omega) = (L-\omega \mathrm{Id})^{-1}$, that is, in terms of $\|R_L(\omega)\|$. As a consequence, the choice of norm (and not only the structure of the operator $L$) plays a critical role in the discussion. In \cite{Jaramillo:2020tuu,Jaramillo:2021tmt,alsheikh:tel-04116011,Gasperin2021} and the following articles, the chosen norm has been the `energy norm' $\|\cdot\||_{{_E}}$ defined in terms of the energy of the propagating linear field. This is a natural norm in many partial differential equation contexts,
in particular to control initial data in second-order Cauchy problems (see also~\cite{Driscoll:1996} in a pseudospectrum related context). In particular, in our BH QNM stability context it seems a very natural norm in the spirit of estimating the `size' of operator perturbations $\delta L$ in terms of the injected energy in the system (although
such relation is far from straightforward, cf.~\cite{Gasperin2021}). However, the fact that the energy norm, in particular the associated `energy scalar product', might be an appropriate one to characterize the size of perturbations from a physical perspective in a `BH QNM stability problem' does not mean that it is also the good one to `define' the studied BH QNMs. This `stability versus definition' problem will be addressed in more detail in \cite{BesBoyJar23}. Here we mention that other norms might, and actually are, key in the BH QNM problem.

Let us mention the following issues regarding the use of the pseudospectrum in the BH QNM stability problem:
\begin{itemize}
\item[i)] {\em Choice of norm.} As discussed above, different problems (e.g. definition vs. stability) may demand different norms leading to distinct pseudospectra. 
\item[ii)] {\em Choice of spacetime slicing.} The geometric character of the BH QNM pseudospectrum, namely the assessment of its independence of the slicing, is still an open problem, in spite of first explorations~\cite{Destounis:2021lum}.
\item[iii)] {\em Matrix approximation to the differential operator.} We compute the pseudospectrum of a matrix approximant $L^N$, not of the actual differential operator $L$,
and this can be the source of issues, e.g. when the actual spectrum of $L$ contains a continuous part in addition to eigenvalues (point spectrum).
\item[iv)] {\em Eigenvalues that are not QNMs.} There may be eigenvalues of $L$ not admitting an interpretation as QNMs~\cite{Warnick:2013hba}. Then, specific assessments are needed.
\item[v)] {\em Instability of spurious eigenvalues.} The use of the approximant $L^N$ may contaminate the pseudospectrum by the instability of eigenvalues of $L^N$, that are not present in the actual spectrum of $L$.
\item[vi)] {\em Convergence of the pseudospectrum.} The status of the convergence of the pseudospectrun of $L^N$ as $N\to\infty$ is an open issue. This is a critical point. 
\end{itemize}

\subsection{Some structural aspects of Schwarzschild-anti-de Sitter geometry}

\subsubsection{Timelike null infinity in AdS}
\label{s:time_null_infinity}
In contrast with the asymptotically flat and dS cases, where the outer (asymptotic) boundary can be naturally chosen as a null hypersurface (null infinity in the former and the cosmological horizon in the latter), null infinity in the
asymptotically AdS case is a timelike surface (cf. e.g. Fig.~\ref{causal_structure}). This makes natural to consider, for asymptotically AdS BHs\footnote{Actually, null foliations can also be used for QNM calculations in the asymptotically flat and dS cases, by imposing vanishing boundary conditions at past null infinity or the past cosmological horizon, respectively~\cite{Jansen:2017oag,Cardoso:2017soq,Cardoso:2018nvb,Destounis:2018qnb}. However, in those cases an actual hyperboloidal slicing is possible and appears as a privileged one.}, two different types of `regular foliations' (not intersecting the horizon bifurcation surface), namely spacelike regular foliations~\cite{Warnick:2013hba,Arean:2023ejh} and null
slicings (e.g.~\cite{ammon2016holographic,chesler2014numerical}).
This provides an assessment to caveat (i) in section VII.A.1 of
\cite{Jaramillo:2020tuu}, namely the dependence of the pseudopectrum on the foliation and therefore its geometric content, in a richer setting than the restricted one implemented in Reissner-Nordstr\"om BHs \cite{Destounis:2021lum}.

In the AdS setting, boundary conditions must be chosen to be
imposed at timelike null infinity, with the crucial requirement of rendering the wave evolution into a well-posed problem. Specifically, the allowed boundary conditions depend (in a given dimension) on the mass of the field. In particular, for the massless (scalar) field that we
will focus on here, only a homogeneous Dirichlet condition is consistent with smoothness 
\cite{friedrich2014ads,holzegel2015einstein}.

\subsubsection{Absence of tails, spectral instability and hydrodynamic modes in AdS}
\label{s:no-tails}
Another crucial aspect of AdS, as it is also the case for dS, is the absence of Price-law tails in the late-time behavior of BH perturbations, corresponding to the absence of a branch cut (or the continuous part of the spectrum in our setting) present in the asymptotically flat case (see also Ref.~\cite{Cardoso:2015fga} for a description on how AdS dynamics connects smoothly to that of asymptotic flat spacetime in the limit of small cosmological constant). Note that the P\"oschl-Teller case, studied in detail in \cite{Jaramillo:2020tuu} as a test-bed for BH QNM (and actually corresponding to scalar field perturbations in dS~\cite{Jaramillo:2020tuu,Bizon:2020qnd}), falls also in this category of potentials without Price-law tails\footnote{See also accelerating Schwarschild and Reissner-Nordstr\"om BHs, described by the $C$-metric, where perturbations decay at late times in an exponential manner \cite{Destounis:2020pjk,Destounis:2020yav}.}.

The use of matrix approximants $L^N$ for the operator $L$,
introduce spurious eigenvalues in the spectrum, corresponding
to a discretization of the `branch cut', whose potential instability contaminates the pseudospectrum. In this context, the AdS case (as well as P\"oschl-Teller and dS) provides a much cleaner BH setting than Schwarzschild or Reissner-Nordstr\"om. This has important consequences at a technical and at a physical level.

At the technical level, in addition to the absence of contamination of the spectra and pseudospectra by spurious eigenvalues in the imaginary axis, the complementary
approach tool to explore the spectral instability by
using random perturbations (`filling' pseudospectra sets, cf. the Bauer-Fike theorem in~\cite{Trefethen:2005,Jaramillo:2020tuu}) can be fully explored in this setting. This was initiated for the P\"oschl-Teller potential in \cite{Jaramillo:2020tuu} (see also~\cite{Sarkar:2023rhp}) and further developed in \cite{Arean:2023ejh} (also \cite{Cownden:2023dam}). Crucially, (random) perturbation analyses are not spoiled by eigenvalues in the `branch cut'.

From the physical point of view, the absence of a non-convergent continuous spectrum along the imaginary axis, permits to study in a clean manner the so-called `hydrodynamic' modes of AdS BHs \cite{Berti:2009kk}, namely QNMs placed also along the imaginary axis corresponding to actual (convergent) eigenvalues, and not to points in the continuum spectrum that manifest as (non-convergent) eigenvalues upon discretisation of the operator. Hydrodynamic modes are relics of holography that connect perturbations of the bulk BH spacetime to corresponding hydrodynamic quantities in the dual CFT at the boundary of AdS \cite{Natsuume:2008ha,Brigante:2007nu,Saremi:2011ab,Delsate:2011qp,Cardoso:2015fga,Berti:2009kk}. These modes can be complex or purely imaginary depending on the perturbation channel of the CFT, are non-trivially connected to the properties of hydrodynamics, and can become long-lived for AdS BHs when the event horizon radius is larger than the AdS length scale \cite{Kovtun:2005ev,Berti:2009kk}. The absence of numerically non-convergent but unstable eigenvalues corresponding to the
`branch-cut' in the continuum limit, will allow us to study the stability of these hydrodynamic  modes, both when they are fundamental, i.e. for `large' Schwarzschild-AdS (SAdS) BHs or when they are overtones, for `small' SAdS BHs. The study of their spectral stability will be a physical aspect on which we will focus in this work.

\subsubsection{QNMs as eigenvalues of a non-selfadjoint operator: the special status of AdS}
\label{s:QNM_eigenvalues}
Independently of the spacetime asymptotics, adopting a spacetime foliation that intersects the BH horizon and $\scri^+$ in regular sections permits to cast the QNM calculation as an eigenvalue problem. For concreteness, let us focus on spacelike regular foliations, leaving the discussion of the setting of QNMs in the null slicing approach for section \ref{s:null_slicing}. Such regular spacelike foliations are referred to as `hyperboloidal slicings' and have been used in the BH QNM setting in the asymptotically flat case (e.g.~\cite{Zenginoglu:2011jz,Ansorg:2016ztf,PanossoMacedo:2018hab,Jaramillo:2020tuu,alsheikh:tel-04116011}) with dS asymptotics (e.g.~\cite{vasy2013microlocal,Sarkar:2023rhp}),
and the AdS case (e.g.~\cite{Warnick:2013hba,Arean:2023ejh}, where they are referred as spacelike `regular slicings'). Considering the wave equation for a scalar field (that can be the one in the master equation of, say, gravitational perturbations) in such hyperboloidal setting, and upon a first-order reduction in time followed by a Fourier transform, the QNM problem can be cast (see section \ref{s:regular_slicing} below) as
\bea
\label{e:QNM_problem}
L u_n = \omega_n u_n \ ,
\eea 
where 
\bea
\label{e:L_operator_intro}
u =
\begin{pmatrix}
  \phi_n \\ \psi_n
\end{pmatrix}  \ \ , \ \ 
L =\frac{1}{i}\!  \left(
  \begin{array}{c|c}
    0 & 1 \\ \hline L_1 & L_2
  \end{array}
  \right) \ ,
  \eea
and operators $L_1$ and $L_2$ are given in Eq. (\ref{e:_L1_L2}) below.

As formulated in Eq. (\ref{e:QNM_problem}), the QNM system looks as a relatively simple and harmless eigenvalue problem, in which even outgoing boundary conditions are no longer explicit, having being incorporated into the bulk of the operator. In addition, the success in \cite{Ansorg:2016ztf,PanossoMacedo:2018hab} where (\ref{e:QNM_problem}) is combined with accurate spectral Chebyshev methods, supports such a vision. However, such apparent simplicity is misleading. In fact, the trade-off of eliminating outgoing boundary conditions in favor of conditions on the regularity of the eigenfunctions introduces a key subtlety in the very definition of QNMs as eigenvalues.

A clear manifestation of such regularity issues is the following key remark in \cite{Ansorg:2016ztf}: considering the convention in which QNM frequencies $\omega_n$ are in the upper-half plane $\mathbb{C}$, it holds that Eq. (\ref{e:QNM_problem}) admits a smooth ($C^\infty$) solution not only for that discrete set $\{\omega_n\}$, but actually for any complex number $\omega$ in the upper-half plane\footnote{The discussion in Sec. IV.A and Appendix B of Ref.~\cite{Ansorg:2016ztf} is presented in terms of the Laplace-transform parameter $s$, rather than the Fourier $\omega$ (related by $s=i\omega$) and without prior first-order reduction in time. Specifically, our Eq. (\ref{e:QNM_problem}) corresponds to Eq. (27) in \cite{Ansorg:2016ztf}, in the particular case of the Schwarzschild BH.}. In other words, any $\omega$ is an eigenvalue for an appropriate smooth eigenfunction. This has the remarkable consequence that initial data can be chosen such that the evolved fields decays as an exponentially damped
sinusoidal time-signal, in which the time decay and the oscillating frequency can be chosen freely (cf. Fig. 3 in \cite{Ansorg:2016ztf} and discussion therein). That is, the field decays exactly in the way QNMs do (and without tails), but for an arbitrary complex frequency. What actually singles out the discrete set of QNM frequencies $\omega_n$ is that the corresponding eigenfunctions possess an enhanced regularity.As discussed in Ref. \cite{Ansorg:2016ztf}, withing the approach originally put forward by Leaver \cite{Leaver85}, this property is cast in terms of the behaviour of the coefficients $H_k$ of a Taylor expansion of the eigenfunction $\phi(\omega)$ around the horizon (a regular singular point): QNMs correspond to $\omega_n$ whose corresponding $\phi_n=\phi(\omega_n)$ present decreasing coefficients $H_k$ as $k\to \infty$, whereas for an arbitrary point $\omega\notin\{\omega_n\}$ coefficients $H_k$ diverge (and still $\phi(\omega)$ is smooth, cf. Appendix B in \cite{Ansorg:2016ztf}).

A systematic manner of dealing with these regularity requirements consists of enforcing QNM eigenfunctions to belong to a Hilbert space where the scalar product is sufficiently stringent to enforce the appropriate regularity. This is the approach taken  by Warnick~\cite{Warnick:2013hba} for the characterization of QNMs as `proper' eigenvalues in asymptotically AdS (as well as dS) BH spacetimes  by using Sobolev's spaces $H^k$ (see also \cite{gannot2014quasinormal,Bizon:2020qnd,ficek2023quasinormal}) and then extended to some BH models in the asymtotically flat case in \cite{Gajic:2019qdd,Gajic:2019oem,galkowski2021outgoing}, in terms of Gevrey-2 classes of functions.

The key point in the setting of the present work is that Warnick's discussion of the asymptotically AdS (and dS) case, in terms of Sobolev's spaces $H^k$ controlling the norm of the first $k$ derivatives\footnote{In this context of control of derivatives, a manifestation of the underlying regularity issues is the observed spectral instability in \cite{Jaramillo:2020tuu,Jaramillo:2021tmt,alsheikh:tel-04116011,Gasperin2021}, where the use of the energy norm (essentially an $H^1$ norm) is not sufficiently stringent to control QNM overtones.}, offers a particularly complete and robust characterization of QNMs as eigenvalues. Unfortunately, such a level of mathematical control has not been yet attained in the generic asymptotically flat case (in terms of so-called Gevrey-2 classes of functions, needing the control of all derivatives). It is for this reason that the AdS case offers a particularly well-suited test-bed for our exploration of the
pseudospectrum and related tools, given the existence of a well-understood mathematical counterpart. The basic elements of QNMs as eigenvalues in AdS will be discussed in Sec. \ref{s:QNM_as_eigenvalues}.

\section{Perturbations of Schwarzschild-Anti de Sitter black holes}
\label{s2}
The SAdS geometry \cite{Griffiths:2009dfa} is given by the line element
\begin{equation}
	{\rm d}s^2=-f(r){\rm d}t^2+f^{-1}(r){\rm d}r^2+r^2{\rm d}\Omega^2,
\end{equation}
where ${\rm d}\Omega^2$ is the line element of the unit sphere, and the redshift function reads
\begin{equation}\label{redshift1}
	f(r)=1-\frac{r_{s}}{r}+\frac{r^2}{R^2},
\end{equation}
with $r_{s}$ and $R$ positive constants. This redshift function has only one zero at a positive value of the radial coordinate, which corresponds to the position of the BH event horizon, designated as $r=r_{h}$ from here on. With this last parameter, the function \eqref{redshift1} can be recast into the form 
\begin{equation}\label{redshift}
	f(r)=\left(1-\frac{r_{h}}{r}\right)\left[1+\alpha^2\left(1+\frac{r}{r_{h}}+\frac{r^2}{r_{h}^2}\right)\right],
\end{equation}
where $\alpha=r_{h}/R$ is a dimensionless constant which relates the size of the BH to the AdS length scale $R$. The form \eqref{redshift1} can be recovered with the relation
\begin{equation*}
	r_{s}=r_{h}(1+\alpha^2).
\end{equation*}
The form \eqref{redshift} will be used throughout this work, since it allows us to simplify the geometric approach of imposing QNM boundary conditions at the horizon.

Assuming linear-field (e.g. scalar or gravitational) perturbations on SAdS spacetime, and imposing appropriate transformations of the degrees of freedom together with a decomposition of the angular sector into spherical harmonics, the radial-temporal part of each angular mode of different types of fluctuations satisfies a wave equation of the form
\begin{equation}\label{waveeq}
	-\phi_{,tt}+\phi_{,r^*r^*}-V_s\phi=0,
\end{equation}
where a comma in the subscript denotes partial differentiation, $r^*$ is the tortoise coordinate, defined as ${\rm d}r^*={\rm d}r/f(r)$, and the potential $V_s$ depends on the angular multipole $\ell$ and the nature of the perturbations. In this work we will mainly focus on scalar $s=0$ and axial gravitational $s=2$ waves, described, respectively, by the potentials~\cite{Berti:2009kk}
\begin{align}
	V_{0}&=\frac{f}{r^2}\left[\ell(\ell+1)+rf'\right],\label{Vsc}\\
	V_{2}&=\frac{f}{r^2}\left[\ell(\ell+1)-\frac{3r_{s}}{r}\right],
\end{align}
respectively, where prime denotes differentiation with respect to the function's argument. For the sake of brevity, in what follows we will present results only for these two potentials. The electromagnetic and polar gravitational cases lead to qualitatively similar results, as we will briefly discuss in the concluding remarks.

\subsection{Boundary conditions}

\begin{figure}
	\centering
	\includegraphics[scale=.8]{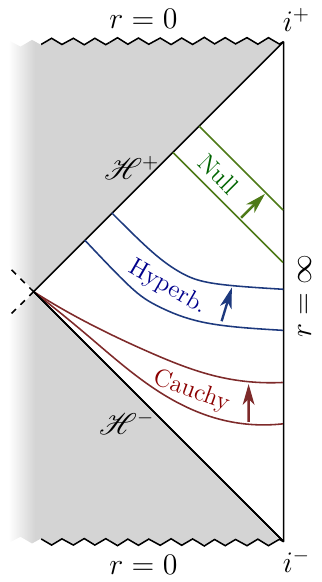}
	\caption{SAdS conformal diagram and different types of hypersurface slicing families on which the evolution of a decaying field outside the horizon can be analyzed.}
	\label{causal_structure}
\end{figure}

Asymptotically AdS spacetimes have a timelike boundary at $r\to\infty$ (see Fig. \ref{causal_structure}), where not just spacelike, but also null and timelike curves can begin and end. Therefore, in order for the evolution of fields on this spacetime to be determined unambiguously from initial data on a spacelike hypersurface, a boundary condition that fits the physical properties of the geometry at $r\to\infty$ must be imposed. A typical choice is the reflective (Dirichlet) boundary condition \cite{Cardoso:2001bb}
\begin{equation}\label{reflective}
	\phi|_{r=\infty}=0,
\end{equation}
which we will adopt in this work. For scalar perturbations $\phi_{sc}$, the divergent potential \eqref{Vsc} at this boundary along with the condition \eqref{reflective} result in a falloff at large $r$ as
\begin{equation}\label{scalarinfty}
	\phi_{sc}\sim\frac{1}{r^2}.
\end{equation}
For the other types of perturbations, the potentials are finite at the boundary, and imposing the behaviour $1/r$ is sufficient for solutions to satisfy the boundary condition.

At the event horizon, all potentials vanish and Eq. \eqref{waveeq} reduces to a free wave equation. The boundary condition there which defines QNMs is that of an ingoing wave,
\begin{equation}\label{ingoing}
	\left(\phi_{,t}-\phi_{,r^*}\right)|_{r=r_{h}}=0.
\end{equation}
A Fourier decomposition on $\phi(t,r_*)$ translates this behavior to the frequency domain as
\begin{equation}\label{singular}
	\phi\sim e^{i\omega(r^*+t)}\sim\left(1-\frac{r_{h}}{r}\right)^{\frac{i\omega r_{h}}{1+3\alpha^2}}e^{i\omega t}.
\end{equation}
From the temporal part we observe that in order for the solution to decay in time, i.e. for the spacetime to be linearly stable, the QNM frequencies $\omega_n$ must have a positive imaginary part. This, however, makes the spatial part having a divergent, oscillatory behavior. This issue is typical of QNMs when working in Schwarzschild-like coordinates, and we will deal with it geometrically in what follows.

\subsubsection{Regular (``hyperboloidal'') slicing}
\label{s:regular_slicing}

Much like how the analytical extension toward the past of BH spacetimes leads to non-physical white hole configurations, the extension of QNMs toward the past describes oscillations with exponentially increasing amplitude, which are equally non-physical. At the horizon bifurcation surface (and at spacelike infinity, if the spacetime were asymptotically flat) this issue is still partially present, leading to the singular oscillatory behavior observed at the boundary in Eq.~\eqref{singular}. In other words, this singular tendency can be seen as a consequence of evolving the wave equation along $t=\text{const.}$ hypersurfaces. These span from the horizon bifurcation surface to the AdS boundary, as shown in Fig.~\ref{causal_structure}, where they are labelled as \textit{Cauchy} slices.

A geometric method of working around this issue is to choose a set of hypersurfaces which intersect the future event horizon $\mathscr{H}^+$, as shown in Fig.~\ref{causal_structure}. One possibility is to use a (half-)hyperboloidal\footnote{We say `half' because the standard hyperboloidal scheme also bends the hypersurfaces toward future null infinity, when this boundary is null.} scheme~\cite{Ansorg:2016ztf,PanossoMacedo:2018hab,PanossoMacedo:2019npm}, which keeps constant-time hypersurfaces spacelike but bends them toward the future horizon. Another is to directly use a foliation of ingoing null hypersurfaces. Both of these transformations, when applied to Eq. \eqref{waveeq}, change the problem in a way that not only makes the QNM solutions regular at the horizon, but also makes regularity equivalent to the ingoing boundary condition \eqref{ingoing}.

Aside from this change in the spacetime foliation, in order to evaluate the problem numerically, we also need to compactify the infinite span of the radial (tortoise) coordinate to a finite domain. We thus define a new coordinate system $\{\tau,\chi\}$ for the radial-temporal sector as
\begin{equation}\label{coords}
	\begin{split}
		t(\tau,\chi)&=\tau-h(\chi),\\
		r^*(\tau,\chi)&=g(\chi).
	\end{split}
\end{equation}
The function $g$ ensures that the $r^*\in(-\infty,r_{\infty}^*)$ interval, where $r^*_{\infty}$ is an integration constant corresponding to $r\to\infty$, is compactified into a finite domain $\chi\in [a,b]$. It is also chosen such that the potential in the wave equations can be written explicitly in terms of $\chi$. A standard choice is the one defined by the relation $\chi=r_{h}/r$, formally given by
\begin{equation}\label{gdef}
	g(\chi)=\int^{r_{h}/\chi}\frac{{\rm d}r}{f(r)}.
\end{equation}
As for the height function $h$, its purpose is to bend the spacetime foliation toward the future horizon $\mathscr{H}^+$, such that the ingoing wave condition is satisfied geometrically, as described above. This is achieved as long as the condition
\begin{equation}\label{hypercond}
	h(\chi)\sim g(\chi)\sim\frac{1}{1+3\alpha^2}\log\left(1-\frac{r_{h}}{r(\chi)}\right),
\end{equation}
is satisfied when approaching the event horizon. In other words, $\tau$ must behave like the Eddington-Finkelstein ingoing null coordinate, $v=t+r^*$, around the horizon. When \eqref{hypercond} is satisfied only in the vicinity of $r_{h}$, while in the rest of the BH exterior the $\tau=\text{const.}$ hypersurfaces are spacelike ($h'<g'$), the slicing is hyperboloidal. In particular, we will work with a slicing
\begin{equation}\label{hhyperb}
	h(\chi)=\frac{1}{1+3\alpha^2}\log(1-\chi),
\end{equation}
obtained by integrating only the dominant term of the redshift function around the horizon in Eq.~\eqref{gdef}, and subsequently checking that the condition $h'<g'$ is satisfied for $r>r_{\rm h}$. Note that for any $\tau$ defined through \eqref{coords}, the Fourier frequency variable $\omega$ is the same as the one corresponding to $t$, since $\partial_\tau=\partial_t$ for these transformations.

After this transformation, and a reduction of order in time through
\begin{equation}\label{order}
	\psi=\partial_\tau\phi,
\end{equation}
the wave equation \eqref{waveeq} can be written as
\begin{equation}\label{wave2}
	L_1\phi+L_2\psi=\partial_\tau\psi,
\end{equation}
where
\begin{equation}
  \label{e:_L1_L2}
	\begin{split}
	  L_1&=\frac{p}{w}\partial_\chi^2+\frac{p'}{w}\partial_\chi-\frac{q}{w} 
          = \frac{1}{w}\left(\partial_\chi(p \; \partial_\chi) - q\right),\\
	  L_2&=2\frac{\gamma}{w}\partial_\chi+\frac{\gamma'}{w}
          = \frac{1}{w}\left(\gamma \partial_\chi + \partial_\chi(\gamma \cdot)\right),
	\end{split}
\end{equation}
with a prime denoting a derivative with respect to $\chi$, and the functions
\begin{equation}\label{wpgq}
	w=\frac{g'^2-h'^2}{|g'|},\quad p=\frac{1}{|g'|},\quad \gamma=\frac{h'}{|g'|},\quad q=|g'|V.
\end{equation}
The first-order problem can be obtained from combining Eqs.~\eqref{order} and~\eqref{wave2},
\begin{equation}\label{wave3}
	i L\,u=\partial_\tau u,
\end{equation}
where we define the vector and block operator (cf. Eq.~(\ref{e:L_operator_intro})) as
\begin{equation}
  \label{e:u_L}
	u=\left(\begin{array}{c}\phi\\ \psi\end{array}\right),\quad L=\frac{1}{i}\left(\begin{array}{c|c}
		0 & 1\\
		\hline
		L_1 & L_2
	\end{array}\right).
\end{equation}

\subsubsection{Null slicing}
\label{s:null_slicing}

The third possibility depicted in Fig.~\ref{causal_structure} is to analyse the wave equation in an ingoing null slicing, with a time parameter given directly by the advanced Eddington-Finkelstein coordinate $v=t+r^*$ (equivalent to a transformation \eqref{coords} with $h=g$). This choice in fact turns out to be the most efficient method for the numerical computations presented in the following sections. We note that a similar approach has been recently pursued in \cite{Cownden:2023dam}. That said, all computations have been performed with both the hyperboloidal and null foliations, for the sake of comparison (as we will see, some of the results differ between the two). In this sense, our work encompasses both the hyperboloidal treatment in Arean et al. \cite{Arean:2023ejh} and the null slicing approach in Cownden et al. \cite{Cownden:2023dam}, allowing us to compare and assess both approches to BH QNM instability in an AdS setting. 

The wave equation \eqref{waveeq} cast in the ingoing null coordinate system acquires the form
\begin{equation}\label{waveeqnull}
	iM\phi=B\partial_v\phi,
\end{equation}
where
\begin{equation}
	\begin{split}
	  M&= \frac{1}{2i}\left[(\partial_\chi p\; \partial_\chi) - q \right] = \frac{1}{2i}\left(p\partial_\chi^2
          +p'\partial_\chi-q\right),\\
		B&=-\partial_\chi,
	\end{split}
\end{equation}
with $p$ and $q$ given again by \eqref{wpgq}.

\subsubsection{AdS boundary}

The ingoing condition at the horizon~\eqref{ingoing} is satisfied automatically for regular solutions in both the hyperboloidal and null coordinates, as can be seen by applying the transformation~\eqref{coords} to~\eqref{ingoing}. On the other hand, the Dirichlet condition, which we have chosen for the AdS boundary, must be imposed by hand. For gravitational (and electromagnetic) perturbations, this condition can be applied directly in the numerical implementation of the problem\footnote{The same method has been used in \cite{Boyanov:2022ark} to apply Dirichlet boundary conditions at the perfectly reflective surface of horizonless compact objects.}, while for scalar perturbations, due to the requirement for a sharper decay in $r$ (see Eq. \eqref{scalarinfty}), an additional transformation must be applied to the equation through the rescaling
\begin{equation}\label{scalarrescale}
	\phi_{sc}(\tau,\chi)=\chi\,\xi(\tau,\chi).
\end{equation}
With this redefinition, Eqs.~\eqref{waveeqnull} and \eqref{wave3} (appropriately regularized with a multiplication by $\chi$ on both sides) automatically impose that $\xi$ must tend to zero at the timelike AdS boundary, which, together with smoothness in $\chi$, completely imposes the required asymptotic behavior~\eqref{scalarinfty}.

However, one important change is worth noting. Eq.~\eqref{wave3}, after the rescaling \eqref{scalarrescale}, takes on the form
\begin{equation}
i\tilde{L}u_\xi=\chi^2\partial_\tau u_\xi,
\end{equation}
with $u_\xi$ a vector of components $\xi$ and $\partial_\tau\xi$, and $\tilde{L}$ obtained by direct substitution of \eqref{scalarrescale} into \eqref{wave3}, followed by a multiplication by $\chi$. The key feature is that the $\chi^2$ multiplying the right-hand side cannot be absorbed into $\tilde{L}$ without making this operator singular at the AdS boundary, due to the very divergence of the potential this transformation is meant to regularise. This makes the problem in the hyperboloidal framework more similar to the one in the null case~\eqref{waveeqnull}, with operators on both sides of the equation. After the Fourier transform discussed below, this equation turns into a \emph{generalized} eigenvalue problem. As for the wave equation in the null case~\eqref{waveeqnull}, after the rescaling~\eqref{scalarrescale} and the multiplication by $\chi$, the right-hand side operator simply turns into $\tilde{B}=-\chi^2\partial_\chi-\chi$.


\section{QNMs as eigenvalues, spectral instability and norm}
\label{s:QNM_as_eigenvalues}

The discussion in the previous Sec.~\ref{s2} leads naturally to casting the calculation of QNMs as an eigenvalue problem in two distinct but (in principle) formally equivalent formulations corresponding, respectively, to the (regular) hyperboloidal and the null slicings.

For the gravitational perturbation problem expressed in the hyperboloidal frame, described in Sec. \ref{s:regular_slicing}, taking a Fourier transform in Eq.~(\ref{wave3}) with respect to $\tau$ leads to the (``frequency domain'') eigenvalue problem for the infinitesimal time generator $L$ in Eq.~(\ref{e:u_L})
\begin{equation}\label{e:eigenvalue_hyper}
L\,u=\omega u.
\end{equation}
On the other hand, when we consider the same problem in the null slicing (\ref{s:null_slicing}), we obtain QNMs characterized in terms of a \emph{generalised} eigenvalue problem\footnote{These generalised eigenvalue problems have been studied in the QNM instability context in \cite{alsheikh:tel-04116011} in optical cavity scenarios.} (see also \cite{Cownden:2023dam}). Specifically, taking the
Fourier transform in the (advanced) time $v$, the wave equation \eqref{waveeqnull} becomes 
\begin{equation}\label{e:eigenvalue_null}
M\phi=\omega B\phi \ .
\end{equation}
Although the analysis of such a generalized eigenvalue problem requires some additional care, a key difference with Eq. (\ref{e:eigenvalue_hyper}) is that such an approach does not involve a previous first-order reduction in time, as in the hyperboloidal approach. This may prove to be relevant in the analysis of the pseudospectrum, as we comment below.

For scalar perturbations, as noted above, the Fourier-transformed equation turns into a generalised eigenvalue problem in both the hyperboloidal and null slicings. In the hyperboloidal one, this is in addition to the time-order reduction, while for the null one the problem remains structurally the same as the gravitational perturbation case \eqref{e:eigenvalue_null}. In the following discussion, the two spectral problems \eqref{e:eigenvalue_hyper} and \eqref{e:eigenvalue_null} will be referenced, and the latter should be understood to encompass the scalar field case for both slicings.

The QNM frequencies, being eigenvalues in spectral problems of the type (\ref{e:eigenvalue_hyper}) and (\ref{e:eigenvalue_null}),
can be defined as $\omega\in \mathbb{C}$ for which the
operators $(L - \omega\mathrm{Id})$ and $(M-\omega B)$, respectively, are not invertible. In other words, frequencies $\omega$ for which there exist eigenfunctions $u(\omega)$ and $\phi(\omega)$ (respectively for Eq. (\ref{e:eigenvalue_hyper})
and (\ref{e:eigenvalue_null})) in the appropriate Hilbert space. Indeed, as discussed in Sec. \ref{s:QNM_eigenvalues}, such $u(\omega)$ and $\phi(\omega)$ QNM eigenfunctions 
must be defined in the appropriate Hilbert space for enforcing
the regularity required to correctly characterize the QNMs.
The choice of Hilbert space is key. In terms of the corresponding resolvent operators, namely 
\bea
\label{e:resolvent_hyper}
R_L(\omega) = (L - \omega\mathrm{Id})^{-1} \ ,
\eea
for the hyperboloidal case and~\cite{Cownden:2023dam,alsheikh:tel-04116011,Trefethen:2005}
\bea
\label{e:resolvent_null}
R_{M,B}(\omega) &=& (M-\omega B)^{-1} \ ,
\eea
for the null slicing, QNM frequencies are those $\omega\in\mathbb{C}$ for which the resolvents $R_L(\omega)$ and $R_{M,B}(\omega)$, defined in the corresponding appropriate Hilbert space, do not exist. In particular, when considering the operator norm $\|\cdot\|$ induced from the
Hilbert space norm, these resolvent norms diverge as $\omega$ tends to one of these eigenvalues.

\subsection{QNMs as eigenvalues in AdS: $H^k$ regularity}
\label{s:QNM_AdS}
As discussed in Sec. \ref{s:QNM_eigenvalues}, the question about the appropriate regularity required to characterize QNMs in asymptotically AdS BHs has been fully addressed by Warnick in~\cite{Warnick:2013hba}. As a result of this analysis, it follows that the upper-complex-half plane $\mathrm{Im}(\omega)>0$ presents a structure in horizontal bands of width given by the BH horizon surface gravity $\kappa$, where increasing regularity of the QNM eigenfunctions is required to identify QNMs as they become more damped. More specifically, focusing on the hyperboloidal slicing eigenvalue problem (\ref{e:eigenvalue_hyper}):

\begin{itemize}
\item[i)] Given $k\in \mathbb{N}$, $H^k$-QNMs are introduced as eigenfunctions of the spectral problem (\ref{e:eigenvalue_hyper}) with $H^k$-regularity, i.e. whose first $k$ derivatives are $L^2$-integrable (namely, they are in the Sobolev space $H^k$). Specifically, defining the $H^k$-regular $L_k$ operator
\bea
\label{e:L_k_operator}
         \begin{array}{ cccc}
         L_k:& H^k\times H^{k-1}&\to & H^k\times H^{k-1} \\ 
         & ( \phi\;\;,\; \psi) & \mapsto & L(\phi,\psi ) = \\
         & & &(\psi, L_1\phi + L_2\psi) 
         \end{array} \ ,
         \eea
$H^k$-QNM frequencies are proper eigenvalues of $L_k$.
\item[ii)] $H^k$-QNM frequencies in the horizontal band above the real axis, characterized by
\bea
  \label{e:AdS_QNM_bands}
  \mathrm{Im}(\omega) < a + \kappa(k -\frac{1}{2}) \ , 
  \eea
form a discrete set (here $a$ is a fixed constant depending on the spacetime and not on $k$).
\item[iii)] The $H^k$-resolvent $R_{L_k}(\omega) = (L_k - \omega\mathrm{Id})^{-1}$, where
\bea
\label{e:H-k_resolvent}
   R_{L_k}(\omega) :  H^k\times H^{k-1}&\to & H^k\times H^{k-1} \ ,
\eea
exists and is a $H^k$-bounded operator in the region defined by inequality (\ref{e:AdS_QNM_bands}), of course except in the discrete set given by $H^k$-QNM frequencies.
\item[iv)] However, for every $k\in \mathbb{N}$, there exists a constant $c_k$ such that $R_{L_k}(\omega)$ does not exist for $\omega\in \mathbb{C}$ with
\bea
\label{e:continuous_k-QNMs}
  \mathrm{Im}(\omega) > c_k \ .
\eea
In other words, all complex numbers $\omega$ in the half-plane defined by (\ref{e:continuous_k-QNMs}) are proper $H^k$-QNM frequencies, namely actual eigenvalues of $L_k$.  
\end{itemize}

In more simple terms, if we want to characterize and locate the QNMs in asymptotically AdS BHs as eigenvalues in the spectral problem (\ref{e:eigenvalue_hyper}), we must proceed band by band, of width $\kappa$, starting from the real axis\footnote{Actually, the starting point depends on the constant $a$ in (\ref{e:AdS_QNM_bands}), but this is just a technical point.} by imposing more and more control on the derivatives of the corresponding eigenfunctions. In order to find the discrete QNMs in the first $\kappa$-band we impose the first derivative to be in $L^2$ (so the eigenfunction is in $H^1$). All frequencies above this band are valid QNMs at this regularity level. Then, to access the discrete QNMs in the second $\kappa$-band we impose second derivatives to be in $L^2$ (eigenfunctions in $H^2$) and so on and so forth: for each $\kappa$-band where we want to locate discrete eigenvalues we must add control on one more derivative. QNMs become more regular as they are more damped, thus requiring more regularity to `pick up' the appropriate discrete QNM frequencies in each new band. But, for any finite $k$, there always remain a continuous set of QNMs with that $H^k$-regularity above the $k$-th $\kappa$-band\footnote{This formalizes in terms of $H^k$ Hilbert spaces, and for asymptotically AdS BHs, the remark in~\cite{Ansorg:2016ztf} commented in Sec.~\ref{s:QNM_eigenvalues}, according to which
all frequencies in the upper complex half-plane are QNMs of the Schwarzschild BH if no appropriate enhanced regularity is enforced on the corresponding eigenfunctions.}.

\subsection{Spectral stability of QNMs: pseudospectrum and choice of norm}
\label{s:Pseudospectrum}
In the previous section we have sketched what in Sec.~\ref{s:Pseudospectrum_open_problems}
we have referred to as the `BH QNM definition problem', in the specific context of asymptotically AdS BHs, where it is fully addressed. We have seen that the choice of the proper Hilbert space, with its associated norm, is key to control the required regularity. The focus of the present manuscript is however placed in the related, but different, `BH QNM stability problem', where the choice of norm is also a key part of the analysis.

\subsubsection{Characterizations of the pseudospectrum}
Non-selfadjoint operators $L$ (and, more generally non-normal operators, for which $[L, L^\dagger]\neq 0$), can potentially suffer from spectral instability, as a consequence of the general non-orthogonality of the eigenfunctions of $L$ and its adjoint $L^\dagger$. In this setting, the spectrum $\sigma(L)$ is not necessarily the proper object to consider, and the notion of $\epsilon$-pseudospectrum $\sigma^\epsilon(L)$ becomes relevant. The latter is defined as the set, for a given $\epsilon>0$, of all complex numbers $\omega\in\mathbb{C}$ that are eigenvalues for `some' perturbed version $L + \delta L$ of the operator $L$, with perturbations $\delta L$ of `size' $\|\delta L\|<\epsilon$ in some appropriate norm. In other words~\cite{Trefethen:2005,Davie07,Sjostrand2019},
\bea
\label{e:pseudo_spectrum_stability}
\sigma^\epsilon(L) = \{\omega\in\mathbb{C}, \exists\delta L, \|\delta L\|< \epsilon:
\omega \in \sigma(L+\delta L)\} \ ,
\eea
Spectrally stable operators are such that the perturbed eigenvalues remain at a distance of order $\epsilon$ under perturbations with $\|\delta L\|\sim \epsilon$, giving rise to $\epsilon$-pseudospectral levels structured as concentric
circles (more specifically `tubular neighborhoods') of radii $\epsilon$. On the contrary, spectrally unstable operators present $\epsilon$-pseudospectra extending over much larger regions in the complex plane. Clearly, such a notion of spectral stability depends on what identifies as a ``big" or ``small" perturbation, and therefore on the choice of norm (see discussion in \cite{Jaramillo:2020tuu,Gasperin2021}). Moreover, the natural choice of norm in this ``stability" problem need not coincide with the appropriate one in the ``definition" problem~\cite{BesBoyJar23}: whereas the latter is dictated by the mathematical structure of the operator, the former may be determined on physical grounds.

An equivalent characterization of the $\epsilon$-pseudospectrum providing a direct link to the resolvent, as well as an efficient manner of determining the $\epsilon$-pseudospectral sets\footnote{Alternatively, the characterization (\ref{e:pseudo_spectrum_stability}) of the $\epsilon$-pseudospectrum can be employed to determine the $\epsilon$-pseudospectra sets by using random perturbations $\delta L$ of norm $\|\delta L\|\leq \epsilon$, cf. \cite{Trefethen:2005}. This has been systematically used in~\cite{Arean:2023ejh} in the AdS black brane setting.} is given in terms of the norm of the resolvent.
In the hyperboloidal case (\ref{e:resolvent_hyper}), this is given by 
\begin{equation}
\label{e:pseudo_spectrum_resolvent_hyper}
\sigma^\epsilon(L) = \{\omega\in\mathbb{C}: \|R_{L}(\omega)\| = \|(L - \omega\mathrm{Id})^{-1}\|>1/\epsilon \},
\end{equation}
whereas in the generalised eigenvalue case (\ref{e:resolvent_null}) we
have\footnote{See Chapter 45 in \cite{Trefethen:2005} for the specifities of pseudospectra in this generalized eigenvalue formulation.}
\begin{equation}
\label{e:pseudo_spectrum_resolvent_null}
\sigma^\epsilon(M,B) = \{\omega\in\mathbb{C}: \|R_{M,B}(\omega)\|= \|(M-\omega B)^{-1}\| >1/\epsilon \}.
\end{equation}
In the case that the norm $\|\cdot\|$ is induced from a scalar
product, the pseudospectrum calculation can be reduced to the
calculation of the minimum of the (generalized) singular values $s^{\mathrm{min}}(A)$ of the appropriate operator $A$ (see details in \cite{Jaramillo:2020tuu}, in particular in its Appendix B). In the hyperboloidal case we have the characterization
\bea
\label{e:pseudo_spectrum_resolvent_hyper_scalarproduct}
\sigma^\epsilon(L) = \{\omega\in\mathbb{C}: s^{\mathrm{min}}(L - \omega\mathrm{Id})<\epsilon \} \ ,
\eea
whereas in the null slicing case, the expression is
\bea
\label{e:pseudo_spectrum_resolvent_null_scalarproduct}
\sigma^\epsilon(M,B) = \{\omega\in\mathbb{C}: s^{\mathrm{min}}(M-\omega B)<\epsilon \} \ .
\eea

\subsubsection{Energy and $H^k$ Sobolev norms}
As we have mentioned above, the natural norm to assess QNM
stability is not necessarily the same as the one needed in order to control the appropriate regularity of the QNMs. Whereas in the asymptotically AdS (and dS) case the latter is given by $H^k$ norms controlling the `size' of the first $k$ derivatives (cf. Sec. \ref{s:QNM_AdS}), from a physical perspective a natural norm concerning the stability issue is
the one related to the energy introduced in the system by the
perturbation $\delta L$. Such an energy norm is essentially an $H^1$ norm, which is natural in the control of the well posedness of the initial value problem of the second-order wave equation (\ref{waveeq}), but it is clearly insufficient to characterize BH QNMs beyond the first $\kappa$-band in the upper complex plane (namely the fundamental or slowest decaying QNM frequency).

Starting from the expression of the energy, in a stationary spacetime, associated with a field $\phi$ in a given hypersurface, we can write explicitly the energy scalar product in the hyperboloidal case, as
(cf. \cite{Jaramillo:2020tuu,Gasperin2021})
\begin{equation}
\label{e:energy scalar product}
\!\!\!\langle u_1,\! u_2\rangle_{_{E}}
\!\!=\!
\frac{1}{2}\!\!\int_a^b \!\!\!\left(\! w(\chi)\bar{\psi}_1 \psi_2 + p(\chi)  \partial_\chi\bar{\phi}_1\!\partial_\chi\phi_2 + q(\chi)\bar{\phi}_1 \phi_2 \!\right)\! d\chi .
\end{equation}
In the null slicing case, the product is essentially the same, keeping in mind that there is no reduction of order in time and, correspondingly, $w(\chi)=0$, leading to
\begin{equation}\label{e:energy scalar product_null}
\langle \phi_1,\! \phi_2\rangle_{_{E}}=\frac{1}{2}\int_a^b \!\left(p(\chi)  \partial_\chi\bar{\phi}_1\!\partial_\chi\phi_2 + q(\chi)\bar{\phi}_1 \phi_2 \!\right)\! d\chi
\end{equation}

For the case of a scalar field, where the function $\phi$ is rescaled according to~\eqref{scalarrescale}, the above expressions apply to the quantity $\chi\xi(\tau,\chi)$. They can be further simplified to remove the cross products between differentiated and non-differentiated functions using integration by parts. Defining the order-reduced rescaled function $\eta(\tau,\chi)=\partial_\tau\xi(\tau,\chi)=\psi(\tau,\chi)/\chi$, and the vector $u_\xi$ with components $\xi$ and $\eta$, the resulting product reads
\begin{equation}\label{e:energy scalar product_scalarf}
\begin{split}
\langle u_{\xi1},\! u_{\xi2}\rangle_{_{E}}&=\frac{1}{2}\int_a^b\left(\chi^2w(\chi)\bar{\eta}_1\eta_2+\chi^2p(\chi)\partial_\chi\bar{\xi}_1\partial_\chi\xi_2\right.\\
&\left.+(\chi^2q(\chi)-\chi p'(\chi))\bar{\xi}_1\xi_2\right)d\chi
\end{split}
\end{equation}
For the null slicing, the above expression once again applies, setting $w(\chi)=0$ and considering the appropriate functions $p(\chi)$ and $q(\chi)$ corresponding to the coordinate transformation.

For higher-order $H^k$ Sobolev norms, we consider an extension to the above expression through the direct addition of a term
\begin{equation}
\chi(1-\chi)\partial_\chi^k\bar{\phi}\partial_\chi^k\phi
\end{equation}
to the integrand. The weight $\chi(1-\chi)$ has been chosen simply in order to avoid adding non-trivial terms at the boundaries, as such terms are directly related to the non-selfadjoint nature of the operator $L$ in the energy norm (see~\cite{Jaramillo:2020tuu}). In practice, any simple weight function, even one which does not vanish at the boundaries, appears to lead to the same qualitative results described below.

Since, both the energy and the $H^k$ norms are induced by the scalar product in the appropriate Hilbert space, we can use the characterizations
(\ref{e:pseudo_spectrum_resolvent_hyper_scalarproduct})
and (\ref{e:pseudo_spectrum_resolvent_null_scalarproduct}) in the standard and the generalised eigenvalue problems, respectively.

\subsection{Numerical implementation}

To obtain the QNM spectrum, and subsequently the pseudospectrum, of the simple and generalised eigenvalue problems discussed above, we follow the method employed in \cite{Jaramillo:2020tuu,Destounis:2021lum,Boyanov:2022ark} and use a Chebyshev-Lobatto grid discretisation in the $\chi$ variable with $N$ grid points. The discretised expressions for both the derivative and integral operators defined above are obtained by expanding the solution in $N$ Chebyshev polynomials, and expressing the derivative/integral of each of them in the same basis, with the operators making the corresponding connection.

The ingoing boundary condition at the horizon is automatically satisfied in the coordinate system used for solutions regular at $r=r_{h}$, which are the only ones the numerical method can approximate. As for the reflective boundary condition at $r\to\infty$, for the axial gravitational perturbations it is imposed by removing a row and column from the discretized matrix operators corresponding to the position of $\chi=0$ in the grid; for scalar perturbations, as mentioned above, the rescaling \eqref{scalarrescale} and the regularisation of the resulting equation already impose the necessary condition for the solutions at the boundary, so no additional steps need to be taken.

In what follows, we use several different values of $N$ for different calculations. For calculating the QNM spectrum of the BHs, we find that $N=100$ is more than sufficient to resolve at least the first 6-7 modes closest to the origin to within several decimal places of accuracy, for all cases studied. For the contour plots of the pseudospectrum, we therefore use the same $N$, finding the result to be qualitatively well behaved. For the QNM spectra of the perturbed operators, we increase the resolution to $N=150$ for the purpose of numerically resolving the highly oscillating perturbation well. Lastly, for the convergence tests we use values in the range $N\in[50,800]$, with jumps of $25$.

For the calculation of the pseudospectrum, we use the generalised singular value decomposition from Eqs.~\eqref{e:pseudo_spectrum_resolvent_hyper_scalarproduct} and \eqref{e:pseudo_spectrum_resolvent_null_scalarproduct}, with
\begin{equation}\label{svd}
s^{\rm min}_E\left(R^{-1}(\omega)\right)={\rm min}\{\sqrt{c}:c\in\sigma((R^{-1}(\omega))^\dagger R^{-1}(\omega))\},
\end{equation}
where $R(\omega)$ stands for the resolvent $R_L(\omega)$ or $R_{M,B}$, defined above for the eigenvalue problems studied, and $(R^{-1}(\omega))^\dagger$ is the adjoint of its inverse. The calculation of the adjoint is performed with respect to the norms descibed above for each case, with its numerical implementation given by
\begin{equation}
A^\dagger=(G)^{-1}\cdot (A^t)^*\cdot G,
\end{equation}
where $A$ is any operator, with its conjugate transpose denoted as $(A^t)^*$, and $G$ is the Gram matrix corresponding to each norm, as defined in Appendix C of \cite{Jaramillo:2020tuu}.

\section{Results}
In this section we present the results concerning the stability of SAdS QNMs, focusing on their generic structural aspects, rather than systematically exploring the parameter space of this problem. We start by presenting the results in the the hyperboloidal slicing setting, and then in the null slicing. After this we discuss and interpret these results in light of Warnick's analytical results in~\cite{Warnick:2013hba}. Finally, this analysis allows us to focus on the physical problem concerning the stability of hydrodynamic modes.

\subsection{Pseudospectra: hyperboloidal slicing}
\label{s:pseudo_hyper}
\begin{figure}
    \centering
    \includegraphics[scale=.6]{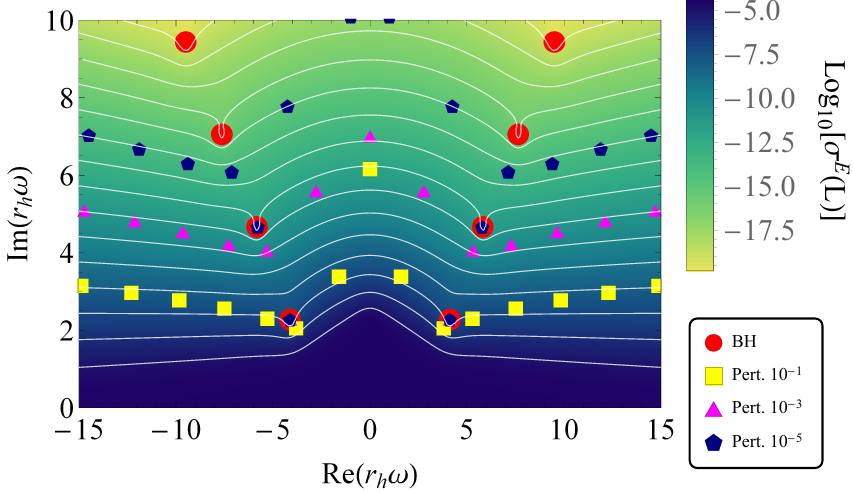}
    \caption{Pseudospectrum in the energy norm of $\ell=2$ scalar perturbations on SAdS with $\alpha=1$, in the hyperboloidal slicing. The spectrum of the unperturbed operator is shown with red circles while the resulting spectra of perturbed versions, where perturbations of the type \eqref{perturbaiton} with $k=20$ have been added to the discretized potential of the operator, are shown with differently shaped and colored points according to their energy norm, indicated in the legend. The grid-point resolution used is $N=100$ for the pseudospectrum and $N=150$ for the perturbed spectra.}
    \label{fsc1_hyper}
\end{figure}

\begin{figure}
    \centering
    \includegraphics[scale=.58]{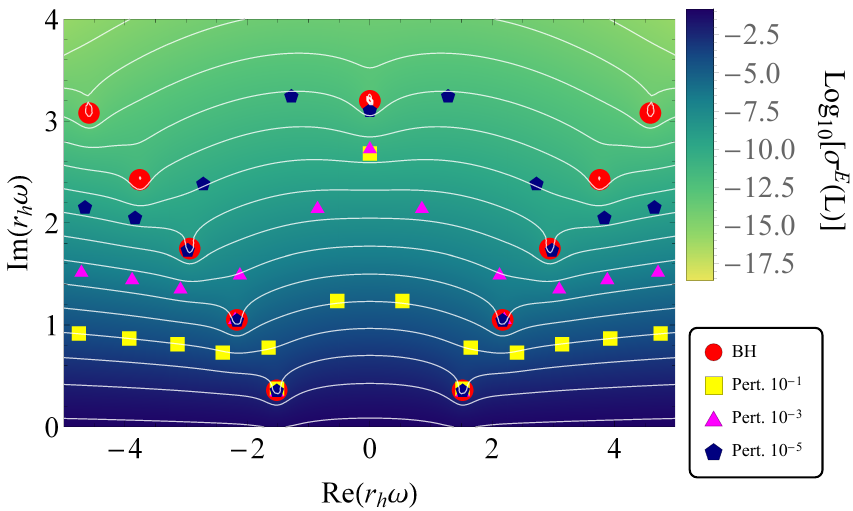}
    \includegraphics[scale=.6]{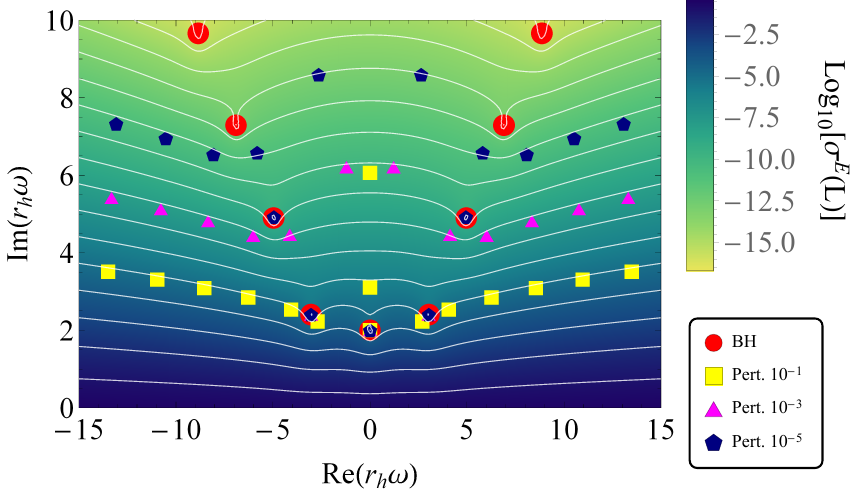}
    \includegraphics[scale=.6]{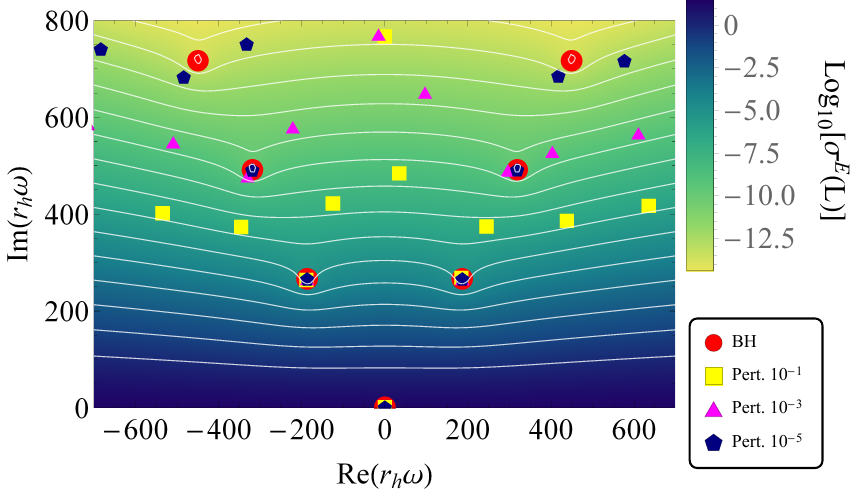}
    \caption{Pseudospectrum of $\ell=2$ axial gravitational perturbations on SAdS with $\alpha=1/2$ (top panel), $\alpha=1$ (middle panel) and $\alpha=10$ (bottom panel), for the hyperboloidal slicing and the energy norm. Again, the spectrum of the unperturbed operator is shown with red circles while the resulting spectrum of perturbed versions of the effective potential under perturbations~\eqref{perturbaiton} with $k=20$ is shown with differently shaped and colored points. Once again, the grid-point number used is $N=100$ for the pseudospectrum and $N=150$ for the perturbed spectra.}
    \label{fax_hyper}
\end{figure}

Figures~\ref{fsc1_hyper} and~\ref{fax_hyper} illustrate, respectively, the stability properties of QNMs for a scalar field and the axial gravitational perturbation (both for the $\ell=2$ mode) in SAdS BHs for given choices of the parameter $\alpha=r_{h}/R$. These and subsequent figures include two independent classes of calculations:
\begin{itemize}
\item[i)] QNM calculations: they corresponds to the resolution (through an appropriate numerical algorithm) of the eigenvalue problem (\ref{e:eigenvalue_hyper}), both for the exact SAdS spacetime (red points) and for effective potentials subject to perturbations $\delta V$ (differently colored and shaped points). These perturbations are of the form
\begin{equation}\label{perturbaiton}
    \delta V=\varepsilon\sin (2\pi k\chi),
\end{equation}
with the wavenumber $k$ generally chosen to be as high as possible while still safely allowing for convergent results with the numerical resolution used (since high wavenumber perturbations lead to the strongest destabilisation, as shown in~\cite{Jaramillo:2020tuu}), and $\varepsilon$ is chosen in a way that fixes the energy norm of the perturbation operator ($\delta L$ or $\delta M$) to the specified values.
\item[ii)] Pseudospectrum calculation: this corresponds to the contour lines map of the function $f_L:\mathbb{C}\to\mathbb{R}^+$, with $f_L(\omega)=\|R_L(\omega)\|$ (actually we plot the inverse $1/f_L(\omega)$) where $R_L(\omega)$ is given in (\ref{e:resolvent_hyper}). Pseudospectral sets $\sigma^\epsilon(L)$ in (\ref{e:pseudo_spectrum_resolvent_hyper}) are given by the
regions bounded below by the $\epsilon$-contour lines.
\end{itemize}

These numerical results recover the qualitative features and contour-line patterns previously found when using the
hyperboloidal slicing to study BH QNM stability in the asymptotically flat
~\cite{Jaramillo:2020tuu,Jaramillo:2021tmt,alsheikh:tel-04116011,Destounis:2021lum,Boyanov:2022ark,Destounis:2023ruj},
asymptotically dS~\cite{Sarkar:2023rhp} and asymptotically AdS~\cite{Arean:2023ejh} configurations.

\subsubsection{Pseudospectra}
\label{s:pseudo_hyper_pseudo}
Concerning the (numerical approximation to the) $\epsilon$-pseudospectral sets, they indicate a strong spectral instability of the BH QNM spectral problem, with increasing instability as $\mathrm{Im}(\omega)$ grows, that is, for higher overtones of the non-perturbed spectrum. The fundamental QNM is the most stable of the QNMs but, at the pseudospectrum level, the transition to instability in the overtones is a smooth one, without any drastic qualitative difference in the transition to the first overtone, as compared to the transition between the first and second  overtones.

Although the pseudospectrum analysis does not allow us to identify the origin of the QNM instability, an interesting point concerns
the asymptotic structure of the contour lines for large $|\mathrm{Re}(\omega)|$. They present a qualitative open structure, already found in other BH settings. More specifically, in the asymptotically flat case (but also in P\"oschl-Teller corresponding to a dS setting~\cite{Bizon:2020qnd}) it was determined~\cite{Jaramillo:2021tmt,alsheikh:tel-04116011,Destounis:2021lum} that the contour lines are consistent with an asymptotic logarithm structure $\mathrm{Im}(\omega)\sim C_1 + C_2\ln(\mathrm{Re} (\omega)+C_3)$, for $|\mathrm{Re}(\omega)|\gg 1$, and support was found for the universality of this structure as long as asymptotic flatness is preserved. 
Remarkably, for asymptotically AdS BHs it was shown in Ref.~\cite{Cownden:2023dam} that this behaviour turns
into a power-law one, namely $\mathrm{Im}(\omega) \sim C_1 +C_2|\mathrm{Re}(\omega)|^\alpha$ for large
$|\mathrm{Re}(\omega)|$ asymptotics.  We have double-checked this original result from Ref.~\cite{Cownden:2023dam} for the cases
corresponding to Figs.~\ref{fsc1_hyper} and~\ref{fax_hyper}, finding agreement with a power-law tendency
(in regions beyond those shown in the plots). From the perspective of the notion of `resonant free region' discussed in~\cite{zworski2017mathematical}, this would indicate that AdS BHs fall into a different `universality class' than asymptotically flat BHs, more akin to the scattering from `impenetrable obstacles' (note the reflecting homogeneous Dirichlet condition at the timelike boundary at infinity) as opposed to the scattering from `(penetrable) potentials' in the asymptotically flat case
(cf. section III.C in~\cite{Destounis:2021lum}). These various behaviours would reflect a non-trivial intertwining between, on the one hand, the underlying `ultraviolet' origin of such asymptotic open branches (since large real frequency $\mathrm{Re}(\omega)$ values explore small-scale structures) and, on the other hand, the `infrared' nature of large scale spacetime asymptotics.

\subsubsection{BH QNM perturbations}
\label{s:pseudo_hyper_QNM_pert}
Regarding the calculation of QNMs as eigenvalues in (\ref{e:eigenvalue_hyper}), QNMs corresponding to the non-perturbed BH potential correctly recover the values found by other methods. In particular, the convergence of such eigenvalues of the $L^N$ finite-rank approximant, with the grid resolution $N$, demonstrates their convergence to the actual QNMs of the differential operator $L$.

Regarding QNM instability, the implementation of perturbations $\delta V$ to $\delta L$ allows us to probe the origin of the instabilities encoded in the pseudospectra figures. In particular, the use of `physical' perturbations $\delta V$ of the effective potential permits us to conclude that high-wavenumber perturbations do trigger QNM instabilities, whereas low-wavenumber perturbations leave the QNM spectrum stable. A crucial point in the asymptotically AdS setting is that the absence of a continuous part of the spectrum along
the upper imaginary axis (corresponding to the branch cut in the scattering resonance approach to QNMs) permits us to use both deterministic and random perturbations of the potentials,
as in the P\"oschl-Teller case in~\cite{Jaramillo:2020tuu}. This is crucial since, on the one hand, this feature permits us to soundly implement a convergence analysis of the perturbed QNMs, namely under deterministic perturbations.
This will be crucial in the convergence issues discussed below.
On the other hand, the possibility of enforcing random perturbations on $V$ (that were not available in the asymptotically flat BH cases due to the contamination from the `branch cut') allows us to push the study of the instability analysis, since they maximize the migration of QNMs to the new Nollert-Price perturbed QNM branches, in particular saturating the pseudospectrum and probing a modified version (namely, a power-law extension) of the (logarithmic) Regge QNM conjecture under `ultraviolet' perturbations~\cite{Jaramillo:2021tmt,Gasperin2021}. While the plots in Figs.~\ref{fsc1_hyper} and \ref{fax_hyper} (and subsequently in Figs.~\ref{fsc1_null} and \ref{fax_null}) only present the QNMs after convergent deterministic perturbations, the same qualitative behaviour has been observed for the case of random perturbations.

The resulting qualitative picture recovers the generic picture of the asymptotically flat case, but in a much cleaner setting due to the absence of `branch cut'. In particular, the fundamental QNM remains much more stable\footnote{Perturbations
non-vanishing at the horizon, not necessarily of `ultraviolet' nature do indeed perturb the fundamental QNM, as stressed in \cite{konoplya2209first} and shown in the AdS context in~\cite{Arean:2023ejh}. In our present context, and in contrast to Ref. \cite{Arean:2023ejh} we are not considering an actual potential in the original geometric wave equation, and the effective one coming from the D'Alembertian do not allow such non-vanishing perturbations at the horizon if we want to preserve the (classical) BH horizon structure.} under such ultraviolet perturbations (but less stable than in asymptotically flat case, as in the dS case\cite{Sarkar:2023rhp}).
Concerning SAdS QNM overtones, they become unstable under high-wavenumber perturbations of the underlying geometry, migrating to perturbed open Nollert-Price-like QNM branches. QNMs in such perturbed Nollert-Price-like branches are stable under further high-wavenumber perturbations. A crucial point to underline here, in view of the latter discussion, is that the exact SAdS QNM overtones are spectrally unstable, and this conclusion can be reached independently of the study of the pseudospectrum. This will prove to crucial in the discussion below.

\subsection{Pseudospectra: null slicing}
\label{s:pseudo_null}
We consider now the BH QNM spectral instability in the setting of null slicings. In particular, together with \cite{Cownden:2023dam}, we present the first BH QNM pseudospectra constructed in a null slicing, since hyperboloidal slicings had been used in the calculation of all previous BH QNM pseudospectra, independently of the $\scri^+$ spacetime null
asymptotics. In accordance with point (ii) in Sec. \ref{s:Pseudospectrum_open_problems}, the calculation of the pseudospectrum in different slicings has the interest of providing a test on the geometric nature (slicing independence) of the pseudospectrum. As we will see below, this comparison problem turns out to be a subtle one, in particular reinforcing the interest of the null slicing pseudospectrum by itself.

\begin{figure}
    \centering
    \includegraphics[scale=.6]{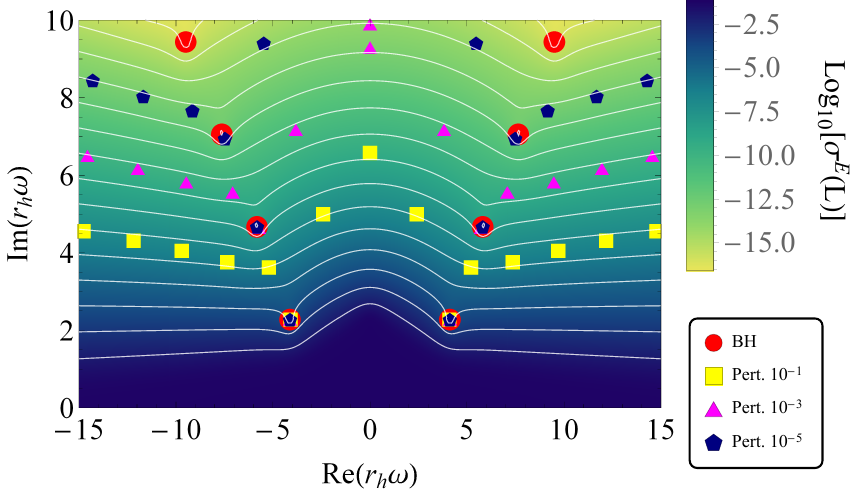}
    \caption{Pseudospectrum in the energy norm of $\ell=2$ scalar perturbations on SAdS with $\alpha=1$, in the null slicing. The parameters and perturbations used are the same as those of Fig.~\ref{fsc1_hyper}.}
    \label{fsc1_null}
\end{figure}

\begin{figure}
    \centering
    \includegraphics[scale=.58]{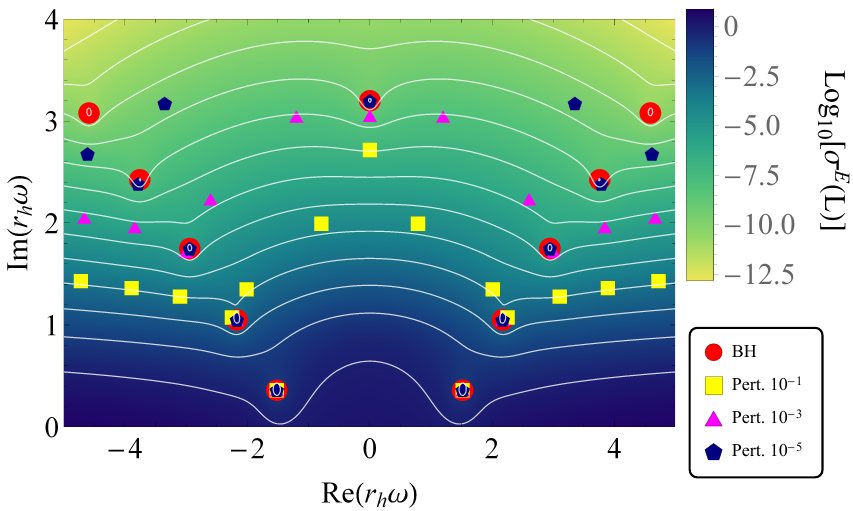}
    \includegraphics[scale=.6]{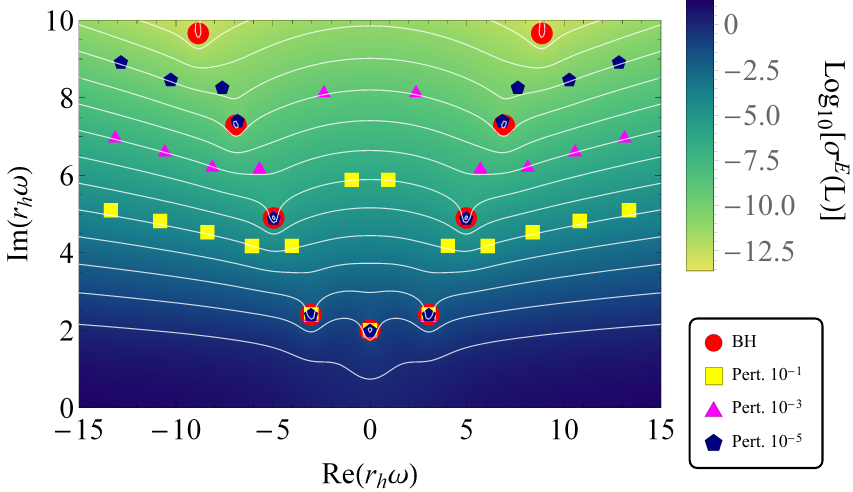}
    \includegraphics[scale=.6]{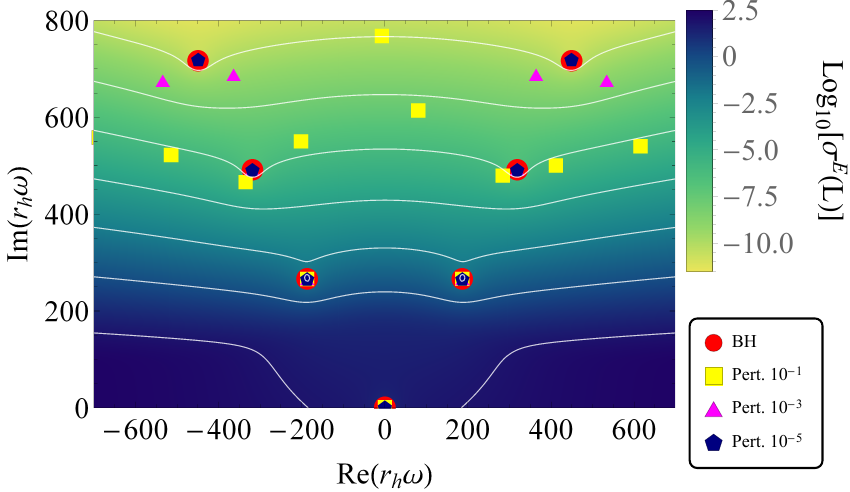}
    \caption{Pseudospectrum of $\ell=2$ axial gravitational perturbations on SAdS with $\alpha=1/2$ (top panel), $\alpha=1$ (middle panel) and $\alpha=10$ (bottom panel). The parameters and perturbations used are the same as those of Fig.~\ref{fax_hyper}.}
    \label{fax_null}
\end{figure}

The spectral stability properties of QNMs of SAdS in the null slicing are illustrated in Figs.~\ref{fsc1_null} and \ref{fax_null}, respectively corresponding to a scalar field and an axial gravitational perturbation (all of the with
$\ell= 2$), and with the same choices for the parameter $\alpha$ as in Figs.~\ref{fsc1_hyper} and \ref{fax_hyper}. As in the hyperboloidal case, such figures present two independent sets of calculations:
\begin{itemize}
\item[i)] QNMs: now calculated from the generalised eigenvalue problem (\ref{e:eigenvalue_null}), for the exact effective potential $V_{\rm eff}=V_s/(\chi^2f)$ of SAdS spacetime (red points) and for perturbed potentials (differently shaped and colored points).
\item[ii)] Pseudospectrum: now giving the `topographic map' of
(the inverse of) function $f_{M,B}:\mathbb{C}\to\mathbb{R}^+$, where $f_{M,B}(\omega)=\|R_{M,B}(\omega)\|$ with $R_L(\omega)$ in (\ref{e:resolvent_hyper}).
\end{itemize}

When considering the qualitative picture of BH QNM instability for overtones in the null slicing, both from the pseudospectra and QNM perturbation perspectives (and with a criterion for `big/small' perturbations given by the energy norm) it coincides with that obtained in the hyperboloidal slicing. Specifically, if we consider the
pseudospectrum far away from the fundamental QNM, we observe in the null slicing (as we did in the hyperboloidal case) an increment of the spectral instability for higher values of $\mathrm{Im}(\omega)$ in the upper-half complex plane, with $\epsilon$-contour lines opening for large $\mathrm{Re}(\omega)$. Likewise, regarding QNM perturbations and as in the hyperboloidal case, QNM overtones migrate to open Nollert-Price-like branches, triggered by high-wavenumber perturbations in the effective potential, whereas the fundamental mode shows a spectrally stable behavior. This provides an overall consistency in that region away the fundamental QNM.

However, discrepancies appear in both the $\epsilon$-pseudospectra contour lines and the perturbed QNMs
as we compare more closely the region around the fundamental QNM. We comment on this below.

\subsubsection{Issues in the comparison of hyperboloidal and null slicing pseudospectra near the fundamental QNM}
\label{s:issues_comparison_pseudospectra}
We comment on two differences in the BH QNM instability in the hyperboloidal and the null slicings:
\begin{itemize}
\item[i)] {\em Pseudospectrum contour lines crossing into the unstable complex half-plane}. The most apparent difference between the hyperboloidal and null slicing pseudospectra is the fact that, in the null slicing case, pseudospectrum contour lines near the fundamental
QNM cross into the lower half of the complex plane. 
Such a behaviour is a potential indication of a transient dynamical phenomenon (see e.g. \cite{Trefethen:1993,Trefethen:2005,Jaramillo:2022kuv,Boyanov:2022ark,Cownden:2023dam}). However this qualitative behaviour of pseudopectrum contour lines is absent in the hyperboloidal case. Actually, this points out a structural difference between the AdS BH QNM pseudospectra in Arean et al. \cite{Arean:2023ejh} (hyperboloidal slicing case) and in Cownden et al. \cite{Cownden:2023dam} (null slicing case). A careful assessment is then needed.

\item[ii)] {\em Calibration of perturbation `sizes' and $\epsilon$-contour values.} When comparing the perturbation of the fundamental QNM in Figs. \ref{fsc1_hyper} and \ref{fax_hyper} (hyperboloidal slicing) with Figs. \ref{fsc1_null} and \ref{fax_null} (null slicing), we observe that the fundamental QNM
seems much more stable in the null slicing case. When trying to assess this issue by using the values in the pseudospectrum contour lines another difficulty arises, namely assigning a neat correspondence between contour lines with the same $\epsilon$ in both sets of pseudospectra.

The origin of these difficulties is two-fold. A first problem concerns the relation between the `size' of the operator perturbation, on the one hand, and the
resulting perturbation of the eigenvalue, on the other side. A neat relation between the module $|\delta \omega|$ of the eigenvalue in the complex plane and the size $\|\delta L\|$ of the operator perturbation holds for the standard eigenvalue problem (\ref{e:eigenvalue_hyper}) and the corresponding $\epsilon$-pseudospectra in Eq. (\ref{e:pseudo_spectrum_resolvent_hyper}) (cf.  \cite{Trefethen:2005}, in particular the discussion around the Bauer-Fike theorem). However, such a relation is more difficult to establish for the generalized eigenvalue problem (\ref{e:eigenvalue_null}). If $B$ were invertible, the proper comparison would be between $|\delta \omega|$ and $\|B^{-1}\delta M\|$ (cf. Chapter 45 in \cite{Trefethen:2005}).
However, in our problem $B$ is not invertible (actually, the corresponding matrices are very ill-posed for inversion). Avenues to circumvent this issue can be explored, 
but the second problem is more serious: the pseudospectrum
in the hyperboloidal slicing does not converge when $N\to \infty$ in the finite-rank approximants $L^N$ to $L$.

\end{itemize}
In the following subsection we discuss this pseudospectrum convergence issue. The AdS structure is key now, since in this case we have a theorem to compare with.

\subsection{Pseudospectra convergence properties: comparison with Warnick's characterization}
\label{s:non-convergence_pseudospectrum}
The non-convergence of the hyperboloidal pseudospectrum with the size $N\times N$ of the matrices $L^N$, signalled at the end of the last subsection, is a major issue in the scheme. Indeed, whereas perturbed eigenvalues in both the hyperboloidal and the null slicing approaches do converge as the numerical grid gets finer, at the pseudospectrum level the situation is less clear, with a convergent behavior in the null slicing case and a non-convergent one in the hyperboloidal setting. QNM spectral instability is not under question, since the convergence of the perturbed eigenvalues soundly demonstrates the ultraviolet instability of QNMs. What is under question is the actual significance of the numerically calculated pseudospectra in the hyperboloidal slicing and, in particular, their capability to `predict' the displacement in the complex plane of QNMs under operator perturbations of `size' $\epsilon$. We address below this pseudospectrum convergence issue, both in the hyperboloidal and the null slicing settings.


\subsubsection{Hyperboloidal slicing}
\label{s:convergence_hyperboloidal}
In order to properly characterize the non-convergence of the pseudospectrum we proceed as follows: (i) we take an arbitrary complex number $\omega$ in the upper-half plane, (ii) we consider a finite rank approximant $L^N$ of the operator $L$ and calculate the norm of its resolvent $\|R_{L^N}\|$, (iii) we take the limit of $\|R_{L^N}\|$ as $N\to\infty$. The statement is then that this limit diverges. In other words, according to our numerical scheme for the pseudospectrum, all points in the upper complex plane are in the spectrum $\sigma(L)$ of $L$.

The last remark is consistent with the statement in \cite{Ansorg:2016ztf}, formulated in the asymptotically flat context, according to which all points in the upper complex plane are eigenvalues of $L$ if we allow their eigenfunctions to be of sufficiently low regularity. In the asymptotically AdS case, however, the results by Warnick~\cite{Warnick:2013hba} provide a sound characterization of the appropriate regularity and the manner to control it in terms of the corresponding Sobolev norm. Indeed, as discussed in Sec. \ref{s:QNM_AdS}, if we consider a band of width $\sim k\cdot\kappa$, the $H^k$-QNM spectrum is a discrete set. In particular, for points $\omega$ in this band that are not in the $H^k$-spectrum, the point (iii) in \ref{s:QNM_AdS} guarantees that the resolvent has finite $H^k$ norm.

\begin{figure*}
\includegraphics[width=8.5cm]{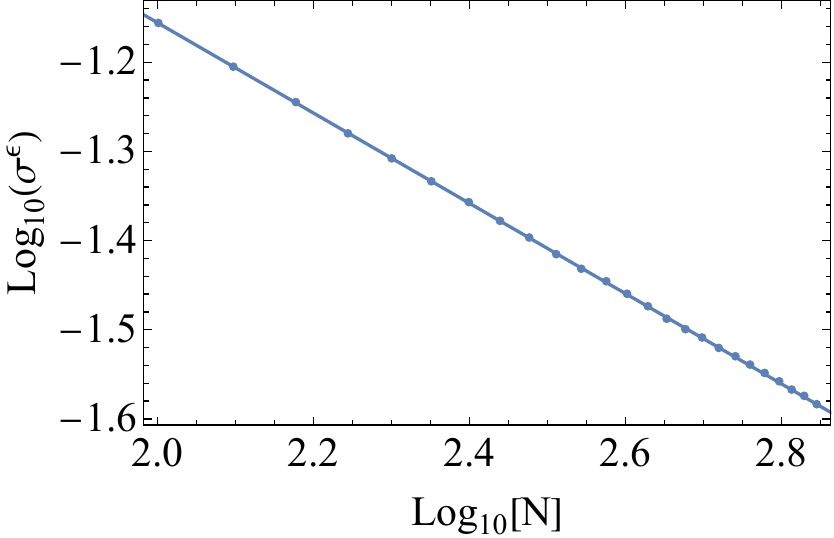}\hspace{0.2cm}
\includegraphics[width=8.5cm]{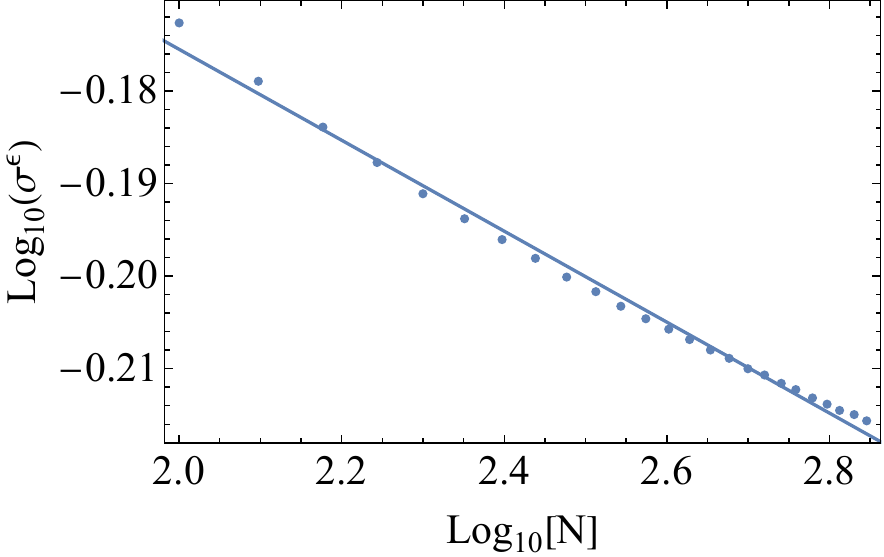} \\
\caption{Norm of (inverse of) $R_L(\omega)$ as function of the numerical resolution $N$, for $\ell=2$ axial gravitational perturbations on an SAdS BH with $\alpha=1$. {\em Left figure}: $H^k, k=1$ (energy norm); $r_h\omega= 2+i0.5$. {\em Right figure}: $H^k, k=1$ (energy norm); $r_h\omega= 2-i0.5$. (These points are not part of the discrete spectrum.) The linear fit in the Log-Log scale shows whether there is an inverse-polynomial approach to zero.
}
\label{fig:non-convergence_hyper}
\end{figure*}

We use the energy norm in our analysis of QNM stability, namely an $H^1$ norm. Therefore, if the resolvent $R_L(\omega)$ of the differential operator $L$ is well attained as the limit of the resolvents $R_{L^N}(\omega)$ of the matrix approximants $L^N$, as $N\to\infty$, then Warnick's theorem implies that we should find convergence in an horizontal band above the real axis with width of the order of the surface gravity $\kappa$, except for the fundamental QNM frequency. However, this is not the case. Indeed, the left panel in Fig.~\ref{fig:non-convergence_hyper} demonstrates a convergence test of the (inverse of the) energy/$H^1$ norm of the resolvent at an illustrative point $\omega=2+i 0.5$, providing strong support for a (power-law) divergence. This behaviour is generic for any point in the upper half plane, no matter how close to the real axis. Even worse, the right panel of Fig.~\ref{fig:non-convergence_hyper} shows the convergence test of the norm of the resolvent for a point in the lower-half plane, showing also non-convergence.

In spite of their unexpected character, these negative results on the convergence of the pseudospectrum (namely of the norm of the resolvent evaluated any point $\omega$) are not in contradiction with Warnick's theorem. The reason is that the resolvent $R_L(\omega)$ is a non-compact operator. As pointed out in (iii) of Sec.~\ref{s:QNM_AdS}, in~\cite{Warnick:2013hba} it has been proven that the resolvent $R_L(\omega)$ exists and it is $H^k$-bounded in the band $\mathrm{Im}(\omega) < a + \kappa(k -1/2)$ (cf. inequality~\ref{e:AdS_QNM_bands}), except in a discrete set of points, namely the QNM frequencies. But it is also shown that $R_L(\omega)$ is not a compact operator. This means that $R_L(\omega)$ cannot be obtained as the limit of a sequence of matrices; in other words, it cannot be approximated (as a whole) by the resolvent of our finite-rank approximants $L^N$.

Assessing the previous point is a subtle issue. On the one hand, we know that partial features of other non-compact operators, such as $L$, admit good approximations by matrices. If this were not the case, we could not approximate the QNMs of $L$ from the eigenvalues of $L^N$. But such a calculation works, because we retain only a subset of all the eigenvalues of $L^N$. However, when calculating the pseudospectrum, the procedure simply fails, and there is no inconsistency in it.  On the other hand, not everything is spurious in the resolvent of finite-rank approximations $L^N$: indeed the open contour-lines of the pseudospectrum of $L^N$ matrices do capture the qualitative distribution of perturbed operators,
that we know to be correct since such perturbed eigenvalues do converge as $N\to\infty$. What fails is the assignment of a particular value $\epsilon$ to a given contour line. In other words, the finite-rank approximants of the non-compact operator $R_L(\omega)$ do capture the qualitative aspects of the pseudospectrum, but fail in the quantitative ones.

\subsubsection{Null slicing}
\label{s:convergence_null}
The situation changes when considering the null slicing. In this case the pseudospectrum presents much better convergence properties, in particular consistent with the horizontal $\kappa$-band structure discussed in section \ref{s:QNM_AdS}. This is in spite of the fact that, a priori, the comparison with the results in~\cite{Warnick:2013hba}, which are presented
in the hyperboloidal setting, is now less straightforward
(in particular regarding the compactness of $R_{MB}(\omega)$). In the following we present results, at an exploratory stage, demonstrating such convergence (for a more detailed study, cf.~\cite{BesBoyJar23}).

As commented above, since the energy norm is an $H^1$ one, we should expect convergence in a horizontal band of approximate width given by the surface gravity $\kappa$ and, then, convergence in this energy norm should fail in all $\omega$'s points above such a band. In order to extend by $\kappa$ the width of the convergent horizontal band in the upper-half plane, we need to control also the $L^2$ size of the second derivatives, i.e. we must  use an $H^2$ norm. Convergence in $H^2$ is then expected in a band of approximate width $2\cdot\kappa$ and, above this band the pseudospectrum in the $H^2$ should not converge.

\begin{figure*}
\includegraphics[width=8.cm]{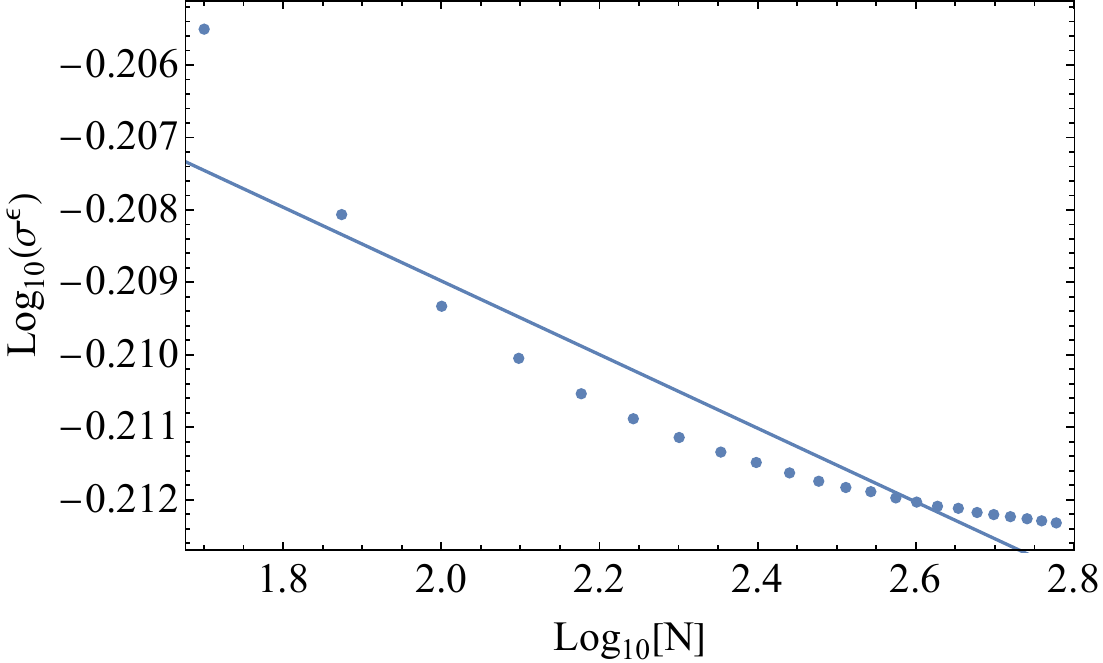}\hspace{0.2cm}
\includegraphics[width=8.cm]{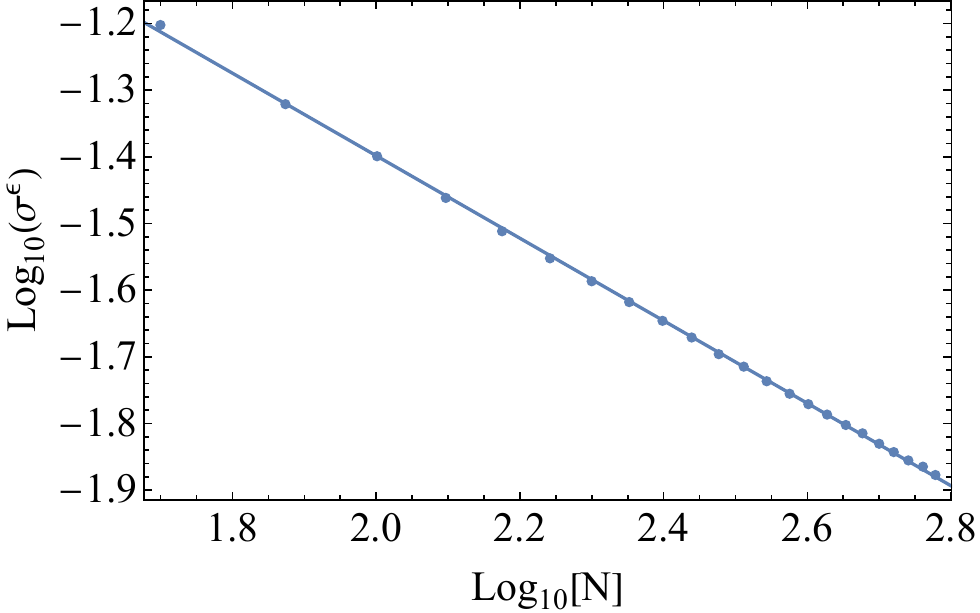} \\
\vspace{0.2cm}
\includegraphics[width=8.cm]{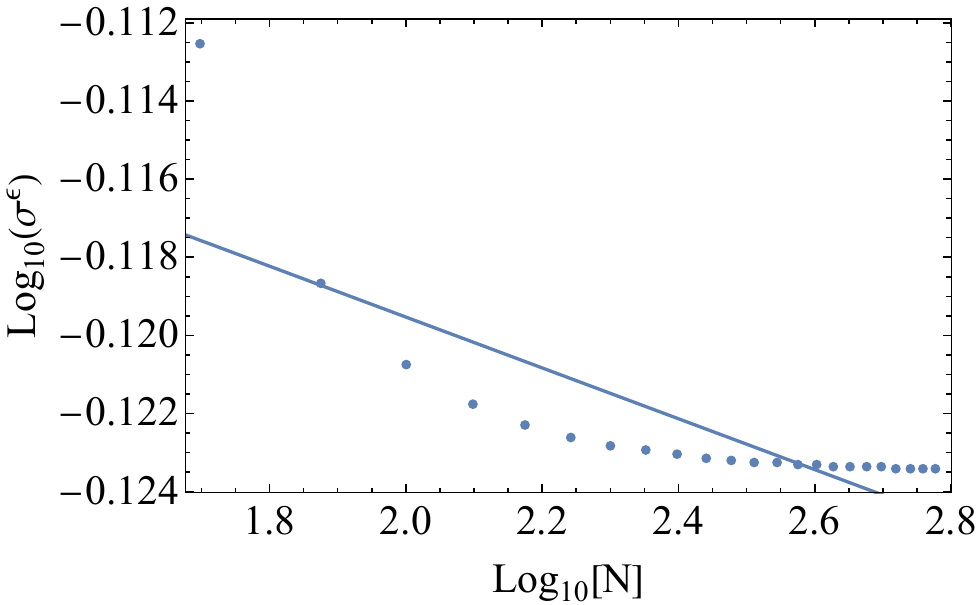}\hspace{0.2cm}
\includegraphics[width=8.cm]{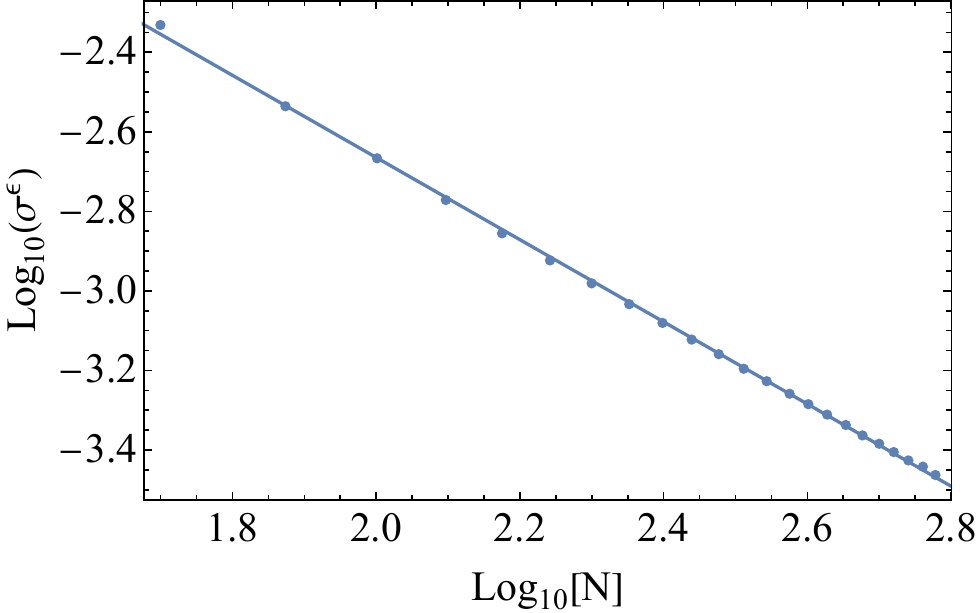} \\
\caption{Norm of (inverse of) $R_{M,B}(\omega)$ as function of the numerical resolution $N$, for $\ell=2$ axial gravitational perturbations on an SAdS BH with $\alpha=1$. The surface gravity of this BH is $\kappa=2/r_h$. {\em Left panel}: first row $H^k, k=1; r_h\omega= 2+i 1.5$, second row $H^k, k=2; r_h\omega = 8+i 3$. {\em Right panel}: first row $H^k, k=1$ (energy norm); $r_h\omega= 2+i 2.6 $, second row $H^k, k=2$; $r_h\omega= 8+i 5$. (These points are not part of the discrete spectrum.) The linear fit in the Log-Log scale shows whether there is an inverse-polynomial approach to zero.
}
\label{fig:convergence-null}
\end{figure*}

In Fig.~\ref{fig:convergence-null} we test this picture (in our setting, we have $\kappa = 2/M$). In the upper panel of the left row, we present the convergence test for the energy norm ($H^1$) resolvent at a point in the first $\kappa$-band, indicating a tendency to convergence in stark contrast with the hyperboloidal foliation. In the upper panel of this first row in Fig.~\ref{fig:convergence-null} we repeat the convergence test in the energy norm, but for a point above the first $\kappa$-band, finding a non-convergent behavior. This is consistent with Warnick's notion of $H^k$-QNMs. In order to further test the latter, in the lower panel of the left row we consider the convergence test in an $H^2$ norm for a point in the second $\kappa$-band, finding a clearly convergent tendency. Finally, keeping the $H^2$ norm but considering a point in the third $\kappa$-band, the lower panel of the right row shows again a non-convergent behaviour. These tests provide a non-trivial confirmation of the $\kappa$-band structure of the SAdS BH, consistent with Warnick's theorem, and illustrate the good convergence properties of the resolvent, and therefore the pseudospectrum, in the null slicing.

The results in this section demonstrate the possibility of using the null slicing for constructing the pseudospectrum of SAdS BHs, in contrast with the hyperboloidal slicing, where convergence issues must be addressed. For this reason, we use the null slicing pseudospectrum to assess in the next section the physical question concerning the spectral stability of SAdS hydrodynamic QNMs.

\subsection{Stability of hydrodynamic modes}
We observe a special mode appearing in the complex plane, namely the hydrodynamic mode. This mode, shown in Fig. \ref{fax_null}, has decreasing imaginary part as the BH radius increases. Approximately when the event horizon radius is comparable to the AdS scale, and beyond, the hydrodynamic mode become the dominant one and approaches the real axis. In this section we investigate its pseudospectral properties, as well as if this mode migrates under perturbations of the effective potential.

\begin{figure}
    \centering
    \includegraphics[scale=.68]{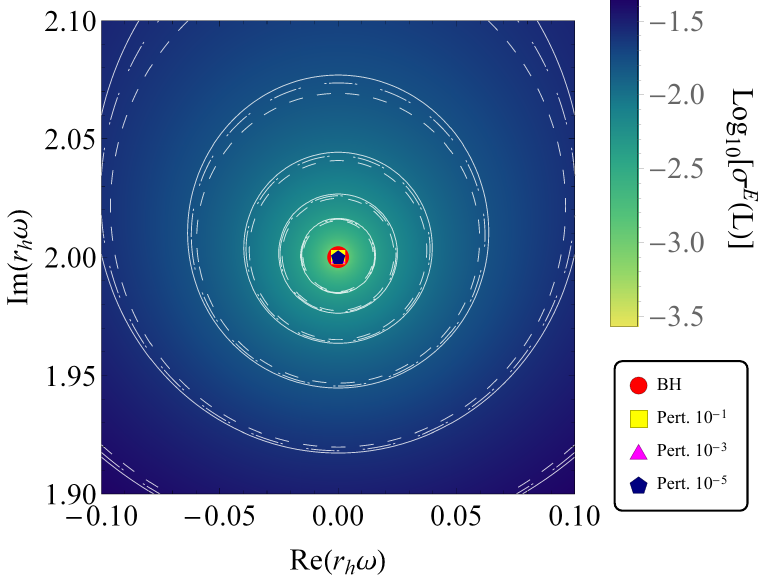}\hspace{0.25cm}
    \includegraphics[scale=.68]{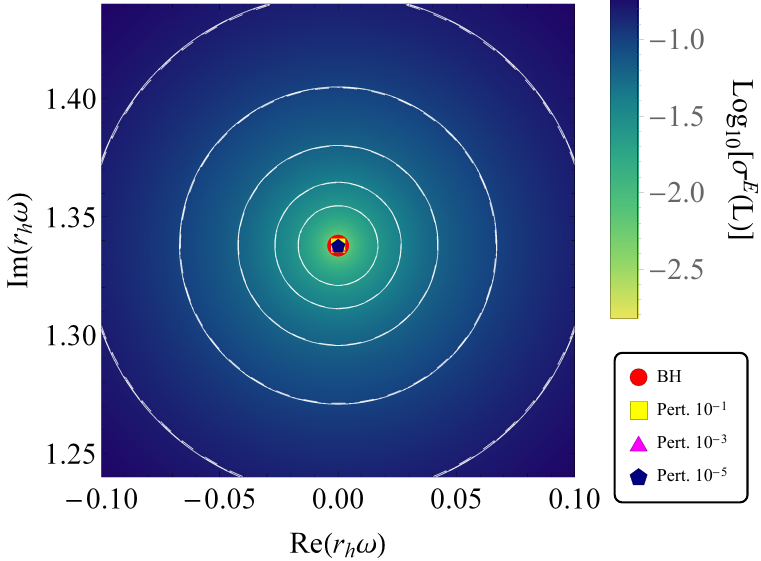}
    \caption{Zoom in around the hydrodynamic mode (shown in red),  for the middle case of Fig.~\ref{fax_null} (top) with $\alpha=1$ and the bottom case of Fig.~\ref{fax_null} (bottom) with $\alpha=10$ panels. The pseudospectrum contour lines are shown for $N=100$ (dashed), $N=150$ (dot-dashed) and $N=200$ (continuous). The differently shaped and colored points designate the perturbed versions of the hydrodynamic modes resulting from random perturbations added to the potential, with energy norm $\|\delta L_1\|\simeq 0.5$.}
    \label{faxz_null}
\end{figure}

The stability of the hydrodynamic mode under the random perturbations considered seems to depend on whether it is the fundamental mode or an overtone. In the case where the hydrodynamic mode dominates the dynamics at late times, it seems to exhibit spectral stability features in contrast to the overtones as it shown in the middle and bottom panels of Fig.~\ref{fax_null}. In contrast, when the hydrodynamic mode is not fundamental then it becomes spectrally unstable as shown in the top panel of Fig.~\ref{fax_null}. Figure \ref{faxz_null} zooms in the pseudospectrum contour lines around this mode for the $\alpha=1$ and $10$ cases, when the mode is spectrally stable, to further enhance our conclusions. The contour lines of rather large values of $\epsilon$ (up to approximately $10^{-1.7}$ in the first case, and up to well above $\epsilon=1$ in the latter case) have a circular shape, which is a direct imprint of spectral stability (see also Fig. 6 in \cite{Jaramillo:2020tuu} and Fig. 10 in \cite{Destounis:2021lum}). Furthermore, the perturbed versions of these modes migrate less than what the pseudospectrum would quantitatively suggest, implying an even higher degree of stability, since we only perturb the effective potential and not the whole operator $L$. To our knowledge, this is the first BH spacetime in the literature that exhibits spectrally stable purely imaginary modes when they are fundamental, while the overtones are still spectrally unstable. 

Despite the qualitative evidence for the stability of the hydrodynamic mode, a quantitative and definitive conclusion relies on a direct identification between the size $\epsilon$ associated to the norm the perturbed term $\|\delta M\|$, with the contour line for the $\epsilon$-pseudospectra $\sigma^{(\epsilon)}$ as defined in Eq.~\eqref{e:pseudo_spectrum_resolvent_null_scalarproduct}, and the circles of radius $\epsilon$ around the mode. Reference \cite{Jaramillo:2020tuu}, however, points out some limitations in the direct identification between the size $\epsilon$ for the operator's perturbation and the contour line for the $\epsilon$-pseudospectra $\sigma^{(\epsilon)}$. From the technical perspective, even though the qualitative behavior for the pseudospectra does not change as one increases the numerical resolution, the $\epsilon$ value associated with a given pseudospectrum contour line does. This lack of absolute convergence for the pseudospectra on the original results raises, at first, a red flag to the conclusion on the spectral stability for the hydrodynamic mode. 

However, we do observe the convergence of the pseudospectra contour lines around the hydrodynamic mode when it becomes the slowest decaying one of the system. Indeed, not only does Fig.~\ref{faxz_null} show that the $\epsilon$-pseudospectral contour lines retain the same values as we increase the numerical resolution (e.g.~$N=100, 150$ and $200$), but also that these lines coincide with the circles of radius $\epsilon$ around the mode under question. The convergence becomes even faster as the hydrodynamic mode moves further apart from the first oscillating mode and approaches the real axis. This result may hint towards an underlying convergence on the pseudospectrum in a larger region of the complex plane as well. 

\section{Conclusions}

This work has focused on assessing some key structural aspects of the spectral stability of BH QNMs, focusing on the convergence properties of the pseudospectrum and using asymptotically AdS BHs as a test-bed, thanks to the existing mathematical characterization by Warnick~\cite{Warnick:2013hba} of QNMs as eigenvalues of non-selfadjoint operators. 

Specifically, we have studied the SAdS BH pseudospectrum in hyperboloidal and null slicings providing, in particular, a common perspective and synthesis of the recent studies of this problem, namely \cite{Arean:2023ejh} in the hyperboloidal slicing and \cite{Cownden:2023dam} in the null case. Our main conclusion is that, in their current (straightforward) numerical implementation, pseudospectrum convergence properties present a better behaviour in the null case and, crucially and non-trivially, are
consistent with Warnick's theorems for AdS BH QNMs. 
It is indeed the existence of such a neat mathematical characterization of BH QNMs in the asymptotically AdS case~\cite{Warnick:2013hba} that has provided the key ingredient allowing us to reach this conclusion on the convergence of the respective numerical schemes. In particular, it has been crucial for demonstrating the structure of the frequency $\omega$-complex stable half plane in horizontal bands of width given by the surface gravity $\kappa$, in such a way that increasing regularity is required to identify QNM frequencies as we explore higher overtones.

Regarding the numerical BH QNM pseudospectra calculated in the hyperboloidal slicing, we observe evidence of non-convergent features in the numerical scheme. Such non-convergence does not hamper the validity of the (asymptotic) structure of the pseudospectrum contour lines, but it critically affects the $\epsilon$ values assigned to such boudaries $\epsilon$-pseudospectra sets. Although the presented analysis has been focused on SAdS BH, the conclusion actually extends to pseudospectra calculated in hyperboloidal slicings for all spacetime asymptotics, namely asymptotically flat, dS and AdS
spacetimes. These poor convergence properties are indeed consistent with the underlying mathematical structure, the ultimate reason being the non-compact nature of the infinitesimal time generator in such hyperboloidal slicings.
Although hints for such non-convergent issues had been found
in the asymptotically flat case~\cite{Jaramillo:2020tuu}, the contamination by spurious eigenvalues in the continuous `branch cut' hindered a clear assessment of this problem. Very importantly, we emphasise that BH QNM spectral instability is not in question, as the convergence of (deterministically) perturbed QNMs demonstrates. The latter had been soundly shown in the asymptotically flat and dS cases and
has now also been established for the AdS case in Refs.~\cite{Arean:2023ejh,Cownden:2023dam} and the present work. Yet, the numerical approximations to the hyperboloidal pseudospectrum must be taken with a grain of salt: although the open contour-lines at a given resolution $N$ do capture the qualitative spectral instability, the non-convergence with $N$  of their associated $\epsilon$ values prevents any quantitative conclusion. In summary, this behaviour signals a fundamental issue of the numerical
scheme for hyperboloidal pseudospectra in its current implementations but, at the same time, it identifies its underlying cause (the non-compactness of the operator) and therefore points to a concrete avenue for its solution. We conclude the urgent need of devoted studies to address, clarify and, hopefully, cure this pseudospectrum non-convergence problem.

Finally, as a concrete application of the insight gained by the  previous discussion on the structural aspects of BH QNMs as eigenvalues of non-selfadjoint operators, we have addressed the physical problem concerning the spectral stability of AdS BH hydrodynamic modes. Specifically, given the good convergence properties of the pseudospectrum in the null slicing, we have chosen this foliation (instead of the 
`worse-posed' hyperboloidal one) to study the problem. This has permitted us to conclude the convergence of such BH AdS hydrodynamic modes. As a by-product, the confidence gained in the null slicing pseudospectrum also provides support for the
potential presence of transients phenomena in the AdS BH setting,
suggested by the crossing of pseudospectra contour lines into the unstable half plane (cf. Fig.\ref{fax_null}). Both the hydrodynamic modes and the possible transients have been discussed in \cite{Cownden:2023dam}, but it is the pseudospectrum convergence analysis here presented and its consistence with Warnick's theorems that provides a solid
picture. 

We plan~\cite{BesBoyJar23} to further push forward these structural studies,  extending the present analysis to include further physical aspects of the problem, that remain open.

\begin{acknowledgments}
We would like to thank Claude Warnick for generously sharing his insights into this problem.
We also thank Daniel Are\'an, J\'er\'emy Besson, Piotr Bizo\'n, Thierry Daud\'e, Dejan Gajic, Jeﬀrey Galkowski, David Garc\'\i a Fari\~na, Jason Joykutty, Karl Landsteiner, Johannes Sj\"ostrand and Maciej Zworski for discussions.
This work was founded by the VILLUM Foundation (grant no. VIL37766), the DNRF Chair program (grant no. DNRF162) by the Danish National Research Foundation, and the European Union’s H2020 ERC Advanced Grant “Black holes: gravitational engines of discovery” grant agreement no. Gravitas–101052587. Views and opinions expressed are however those of the author only and do not necessarily reflect those of the European Union or the European Research Council. Neither the European Union nor the granting authority can be held responsible for them.
V.B. acknowledges support from the Spanish Government through the projects PID2020-118159GB-C43, PID2020-118159GB-C44 (with FEDER contribution).
V.C.\ is a Villum Investigator and a DNRF Chair. 
This project has received funding from the European Union's Horizon 2020 research and innovation programme under the Marie Sklodowska-Curie grant agreement No 101007855.
K.D. acknowledges financial support provided under the European Union’s H2020 ERC, Starting Grant agreement no. DarkGRA–757480 and the MIUR PRIN and FARE programmes (GW-NEXT, CUP:B84I20000100001). K.D. also acknowledges hospitality and financial support provided by the Theoretical Astrophysics group of the University of T\"ubingen and Instituto Superior T\'ecnico where parts of this project were worked out.
J.L.J. acknowledges support from the EIPHI Graduate School (contract ANR-17-EURE-0002), 
the ANR ``Quantum Fields interacting with Geometry” (QFG) project
(ANR-20-CE40-0018- 02), and from the Spanish Government through the projects PID2020-118159GB-C43/AEI/10.13039/501100011033 and by the Junta de Andaluc\'\i a through the project FQM219, as well as financial support from the Severo Ochoa grant CEX2021-001131-S funded by MCIN/AEI/ 10.13039/501100011033.
R.P.M. acknowledges the financial support provided partially by STFC via grant number ST/V000551/1 STFC via grant number ST/V000551/1.
We thank FCT for financial support through Project~No.~UIDB/00099/2020.
We acknowledge financial support provided by FCT/Portugal through grants PTDC/MAT-APL/30043/2017 and PTDC/FIS-AST/7002/2020.
This research project was conducted using the computational resources of ``Baltasar Sete-Sois'' cluster at Instituto Superior T\'ecnico.
\end{acknowledgments}

\bibliography{AdS_pseudobib}

\end{document}